\documentclass[journal]{IEEEtran}
\ifCLASSINFOpdf
\else
\fi
\usepackage{graphicx}          
\usepackage[dvips]{epsfig}    
\usepackage{color}
\usepackage{xcolor}

\usepackage{graphicx}
\usepackage{epstopdf}

\usepackage{amsfonts}
\usepackage{amsmath}

\usepackage{caption2}
\usepackage{enumerate}

\usepackage{algorithm}
\usepackage{algorithmic}
\usepackage{booktabs}
\usepackage{setspace}

\usepackage[T1]{fontenc}
\usepackage{stfloats}

\usepackage{url}
\def\QEDclosed{\mbox{\rule[0pt]{1.3ex}{1.3ex}}} 

\usepackage{ntheorem}
\theoremheaderfont {\itshape}
\theoremseparator {:}
\theorembodyfont{\normalfont}

\newtheorem*{pf}{\indent Proof}
\newtheorem{rem}{\indent Remark}
\newtheorem{thm}{\indent Theorem}
\newtheorem{lem}{\indent Lemma}

\newtheorem*{nota}{\indent Notation}

\definecolor{V1}{rgb}{0,0,0}
\definecolor{V2}{rgb}{0,0,0}
\definecolor{V3}{rgb}{0,0,}
\definecolor{V3-2}{rgb}{0,0,0}

\makeatletter
\long\def\@makecaption#1#2{%
	\vskip\abovecaptionskip
	\sbox\@tempboxa{#1: #2}%
	\ifdim \wd\@tempboxa >\hsize
	#1: #2\par
	\else
	\global \@minipagefalse
	\hb@xt@\hsize{\box\@tempboxa\hfil}%
	\fi
	\vskip\belowcaptionskip}
\makeatother

\begin{document}
%
\title{\textcolor[rgb]{0,0.267,0.588}{Learning Hidden Markov Models for Linear Gaussian Systems with Applications to Event-based State Estimation}}
%
%
%

\author{Kaikai~Zheng,~Dawei~Shi, and~Ling~Shi%
\thanks{K. Zheng and D. Shi are with the State Key Laboratory of Intelligent Control and
	Decision of Complex Systems, School of Automation, Beijing Institute of
	Technology, Beijing 100081, China. (e-mails: KaiKai.Zheng@bit.edu.cn, daweishi@bit.edu.cn)}%
\thanks{L. Shi is with the Department of Electronic and Computer Engineering,
	Hong Kong University of Science and Technology, Kowloon,
	Hong Kong (e-mail: eesling@ust.hk).}}%

%
%

\markboth{Journal of \LaTeX\ Class Files,~Vol.~xx, No.~x, April 2020}%
{Shell \MakeLowercase{\textit{et al.}}: Bare Demo of IEEEtran.cls for IEEE Journals}
%



\maketitle
\textcolor[rgb]{0.157,0.373,0.302}{\begin{abstract}
\textcolor{black}{{\color{V2}This work attempts to approximate a linear Gaussian system with a finite-state hidden Markov model (HMM), which is found useful in solving sophisticated event-based state estimation problems.} An indirect modeling approach is developed, wherein a state space model (SSM) is firstly identified for a Gaussian system and the SSM is then used as an emulator for learning an HMM. In the proposed method, the training data for the HMM are obtained from the data generated by the SSM through building a quantization mapping. Parameter learning algorithms are designed to learn the parameters of the HMM, through exploiting the periodical structural characteristics of the HMM. The convergence and asymptotic properties of the proposed algorithms are analyzed. The HMM learned using the proposed algorithms is applied to event-triggered state estimation, and numerical results on model learning and state estimation demonstrate the validity of the proposed algorithms.  }
\end{abstract}}

\textcolor[rgb]{0.157,0.373,0.302}{\begin{IEEEkeywords}
\textcolor{black}{Event-based state estimation, linear Gaussian system, hidden Markov models, parameter learning.}
\end{IEEEkeywords}}

%
\IEEEpeerreviewmaketitle

\textcolor[rgb]{0,0.267,0.588}{\section{\textsc{Introduction}}}
\IEEEPARstart{T}{he} widespread utilization of wireless communication in civil and industrial applications promotes the development of networked control systems \cite{98}. {\color{V3-2}The requirement on system performance under limited communication bandwidth and computation leads to the development of event-based control and estimation strategies \cite{aastrom2002comparison,hjt12}, with which network resources can be saved significantly while maintaining system performance.}

An important aspect of event-based networked control system design is state estimation with event-triggered sensor measurements \cite{ib05, rmb12,llw10,ms10,mh12,shi2016stochastic}. A number of interesting attempts were made in the literature, mostly for linear Gaussian systems. For reliable channels, systematic approaches were proposed to schedule the networked sensors or to design the event-based state estimators. For instance, in \cite{jqkevent}, a minimum mean square error estimator was derived for an event-triggering scheme to quantify the magnitude of the innovation of the estimator, which was achieved based on Gaussian assumptions {\color{V3}of the distributions of the states conditioned on} the available hybrid measurement information. More general event-triggering conditions and multiple sensor measurements were discussed in \cite{SHI20141641}. However, for unreliable channels, the authors in \cite{ej2017,8920146} proved that one cannot find an event-based trigger such that under possible packet drops, the state variable could remain a Gaussian variable. Thus, the Gaussian filtering framework might not be appropriate for event-triggered state estimation problems with unreliable communication channels.

To deal with such  problems, event-triggered state estimators based on hidden Markov model (HMM) received much attention during the past few years \cite{6906271,Shi2016Event, ref15, ref16}. For instance,  a Markov chain approximation algorithm for event-based state estimation was   presented in \cite{6906271}. In \cite{Shi2016Event,ref15}, the reference measure approach was shown effective in obtaining  recursive event-based state estimates for  HMMs with different lossy channels.  However, modeling a dynamic system as an HMM is a necessary task to implement the estimators aforementioned, which motivates our research in this work. 

HMM is a data-driven generative model {\color{V3}whose} mathematical foundations were set off in the 1960s \cite{Baum1966Statistical} and it was widely applied in many fields (e.g., natural language processing \cite{YUYAN2016,YUYAN2017}, image processing \cite{Image20171,Image20172,Image2016}, pattern recognition \cite{P2017,P2016,P2016j} and so on). To model a system as an HMM, different approaches have been attempted. First, several methods were  developed to choose states that described system dynamics. The authors in \cite{hmm1992nor} designed a probabilistic classification method for HMM states which made the HMM more robust against the possible mismatch or variation between training and test data. {\color{V3-2}Algorithms to quantize continuous vectors to finite discrete states were discussed in \cite{Huang1989Semi,Ilyas2008Speaker}.} Second, sampling strategies were also investigated. Monte Carlo Markov Chain is a typical sampling method including Gibbs sampling, Metropolis-Hastings sampling and so on \cite{oro22547}. A sampling algorithm that resampled the full state sequence by employing a truncated approximation of Dirichlet process was developed to improve the mixing rate \cite{inproceedingsfox}. Algorithms were proposed for the problem of generating a random sample in accordance with the steady-state probability law of Markov chains \cite{Propp1998How}. Finally, after choosing states and generating sample data, algorithms for estimating the parameters of an HMM were also {\color{V3}developed}. On one hand, algorithms based on expectation maximization (EM) were designed to calculate the maximum likelihood estimation of the parameters of an HMM on-line \cite{Collings1994On,Krishnamurthy2002Recursive} or off-line \cite{Ghahramani1997Factorial,MCH2010}. On the other hand, some works \cite{AN1998,zyc2009} developed alternative estimation procedures named corrective training and maximum mutual information estimation, which were designed to minimize the recognition errors instead of minimizing the estimation errors of {\color{V3}the} parameters. Besides, realization problems were discussed in \cite{Anderson1999The,HuangMinimal} to establish an HMM given observation sequence only.

Compared with the existing methods aforementioned, modeling a dynamic system with continuous states as an HMM involves additional challenges. The choice of states will affect the complexity of the HMM. {\color{V3}Although the algorithms for choosing the states of an HMM  in \cite{Huang1989Semi} and \cite{Ilyas2008Speaker} can be applied to linear Gaussian systems, these methods are complicated whose complexity increases in the order of the systems and the dimension of the states.} In this work, a simpler mapping is constructed between the quantized values of  $n$-dimensional vectors and the states of an HMM, based on which we can describe higher-order dynamic systems with the HMM states. Although the HMM has been extensively applied to one-dimensional problems, the extension to higher dimensions is intractable in many cases because of the data size and computational complexity \cite{Perronnin2003Iterative}. To overcome this difficulty, the structure and the pattern of transition matrices of linear Gaussian systems are investigated in this work, which are used to simplify our model training. Furthermore, in the aforementioned works \cite{inproceedings,AN1998,inproceedings2,article3,article4,article5}, an HMM was trained by data sampled from real systems directly but it is difficult to sample dynamic systems with the same methods. Based on this observation, we develop an indirect approach through building an emulator leveraging existing results in system identification. The main contributions of our work are summarized as follows:

\begin{enumerate}[1)]
	\item An indirect approach of training an HMM is proposed, in which we model a dynamic system as a state space model (SSM) first and then learn an HMM by using data generated by the SSM.  
 Specifically, a mapping is constructed between quantized $n$-dimensional state space of the SSM and the discrete states of an HMM. Based on the proposed mapping, the SSM is used as an emulator that generates the state and measurement information needed to learn the HMM. 
	\item Patterns and characteristics of an HMM of a Gaussian system are investigated and used to simplify our HMM parameter learning algorithms. The core idea comes from the  similarity in the probability density functions of different blocks in the transition matrix of an HMM. A series of algorithms are designed to reduce the time complexity of learning an HMM. Upper bounds of parameter estimation errors of transition matrices are proposed and we show that the error bounds diminish to zero with the increase of the quantization precision.
	\item The obtained HMM learning approach is applied to event-based state estimation with an unreliable communication channel. Through a numerical example with comparative results from a Kalman filter with intermittent observations {\color{V3}\cite{bs2004}}, we show that the HMM-based estimator performs nearly as {\color{V3}good} as the Kalman filter when {\color{V3}the} communication rate is high, but performs better than the Kalman filter with the reduction of {\color{V3}the} communication rate, which {\color{V3}demonstrates} the validity and strength of the proposed learning approach.
\end{enumerate}

The {\color{V3}remainder} of the paper is organized as follows. Section II presents the system description and problem formulation. An indirect method for modeling a Gaussian system as an HMM is provided in Section III. Implementation issues and numerical verification of the results are presented in Section IV, followed by some conclusions in Section V.

\begin{nota} For a matrix $A=(a_{i,j})\in\mathbb{R}^{p\times q}$, let $[A]_i=\left[a_{i,1},{\color{V3}\ldots},a_{i,q}\right]$ and $[A]^j=\left[a_{1,j},{\color{V3}\ldots},a_{p,j}\right]^{\mathrm{T}}$. We use $x_i$ to denote the $i$th element of vector $x$; thus for   $x\in\mathbb{R}^n$, we write $x=\left[x_1, x_2, {\color{V3}\ldots}, x_n\right]^{\mathrm{T}}$. Define ${\rm sum}(x):=\sum_{i=1}^n x_i$. Write the Euclidean norm of real vectors as $\|\cdot \|_2$ and the absolute value of a scalar as $|\cdot |$. The notation $\lfloor x\rfloor$ denotes the flooring operator for a scalar $x$. The notation $\delta_N^i$ is used to denote the $i$th column of $I_N$, where $I_N\in\mathbb{R}^{N\times N }$ is an identity matrix. A diagonal matrix with elements of the main diagonal equal to $\{q_1,{\color{V3}\ldots},q_n\}$ is written as $Q={\rm diag}\{q_1,{\color{V3}\ldots},q_n\}$. Kronecker product and Khatri-Rao product are denoted by  $\otimes$ and $*$, respectively. Let $A=(a_{i,j})\in\mathbb{R}^{m\times n}$, $B=(b_{i,j})\in\mathbb{R}^{p\times q}$ and $C=(c_{i,j})\in\mathbb{R}^{o\times n}$, the Kronecker product and the Khatri-Rao product are defined as
	\begin{align*}
	A\otimes B &:=\left[\begin{array}{cccc}
	a_{1,1}B & a_{1,2}B & \cdots & a_{1,n}B \\
	a_{2,1}B & a_{2,2}B & \cdots & a_{2,n}B \\
	\vdots & \vdots & \ddots & \vdots \\
	a_{m,1}B & a_{m,2}B & \cdots & a_{m,n}B
	\end{array}\right], \\
	A*C &:=\left[[A]^1\otimes[C]^1, {\color{V3}\ldots}, [A]^n\otimes [C]^n\right].
	\end{align*}
{\color{V2}In the derivations, the notions of `standard column'  and `target column' are used.  Variables for `standard column' are marked by `${\rm\textsc s}$' and variables for `target column' are marked by `${\rm \textsc t}$'. For example, $\mu_{\rm\textsc s}^x$, $x_p(k;{\rm\textsc s})$ are variables related to `standard column', while $\mu_{\rm\textsc t}^x$, $x_p(k;{\rm\textsc t})$ are variables related to `target column'.}
\end{nota}

\textcolor[rgb]{0,0.267,0.588}{\section{\textsc{Problem Setup}}}
Consider an $n$th-order linear Gaussian system with the following state space model 
 {\color{V3}\begin{equation}\label{equ:sys_without input}
\begin{split}
x(k+1)&={\color{V3}\mathcal{A}}x(k)+w(k), \\
y(k) &={\color{V3}\mathcal{C}}x(k)+v(k),
\end{split}
\end{equation}}where ${\rm x}(k)=\left[{\rm x}_1(k),{\color{V3}\ldots},{\rm x}_n(k)\right]^\mathrm{T}\in\mathbb{R}^n$ is the state, ${\rm y}(k)=\left[{\rm y}_1(k),{\color{V3}\ldots},{\rm y}_m(k)\right]^\mathrm{T}\in\mathbb{R}^m$ is the output, $w(k)\in\mathbb{R}^n$ and $v(k)\in\mathbb{R}^m$ are i.i.d. zero mean Gaussian noises with covariance $Q={\rm diag}\{q_1^2,{\color{V3}\ldots},q_n^2\}$ and $R={\rm diag}\{r_1^2,{\color{V3}\ldots},r_n^2\}$. For learning purpose, we assume the system is stable. Note that although we require the covariance matrices $Q$ and $R$ to be diagonal, the case of non-diagonal $Q$ and $R$ can be considered through performing linear transformations of the state and measurement equations in {\color{V3-2}\eqref{equ:sys_without input}}. In this work, we attempt to  build an HMM representation for the system in {\color{V3-2}\eqref{equ:sys_without input}}. To aid our description, the notation related to an HMM is first introduced.

The state space $S_X$ and the measurement space $S_Y$ of an HMM are defined as finite sets
\begin{equation}\label{equ:sets}
S_X:=\left\{\delta_N^1,{\color{V3}\ldots},\delta_N^N\right\},\quad S_Y:=\left\{\delta_M^1,{\color{V3}\ldots},\delta_M^M\right\},
\end{equation}
where the positive constants $N$ and $M$ denote the cardinalities of $S_X$ and $S_Y$, respectively. Let $X$ be a Markov chain with state $X(k)\in S_X$ and the Markov property implies that
\begin{align*}
&{\rm P}\left(X(k+1)=\delta_N^j|\{X(0),{\color{V3}\ldots},X(k)\}\right)\\
=&{\rm P}\left(X(k+1)=\delta_N^j|X(k)\right).
\end{align*}
{\color{V3}Write
\begin{align*}
a_{i,j}&:={\rm P}\left(X(k+1)=\delta_N^i|X(k)=\delta_N^j\right),\\ A&:=(a_{i,j})\in\mathbb{R}^{N\times N},
\end{align*}
and we obtain
\begin{equation}\label{equ:A}
E(X(k+1)|X(k))=AX(k).
\end{equation}
The measurement process satisfies
\begin{align*}
&{\rm P}\left(Y(k)=\delta_M^j|\{X(0),{\color{V3}\ldots},X(k)\}\right)\\
=&{\rm P}\left(Y(k)=\delta_M^j|X(k)\right).
\end{align*}
Similarly, after defining
\begin{align*}
c_{i,j}&:={\rm P}\left(Y(k)=\delta_M^i|X(k)=\delta_N^j\right),\\ C&:=(c_{i,j})\in\mathbb{R}^{M\times N},
\end{align*}
we obtain}
\begin{equation}\label{equ:C}
E(Y(k)|X(k))=CX(k).
\end{equation}
\begin{figure}[t]
	\flushleft
	\includegraphics[width=\linewidth]{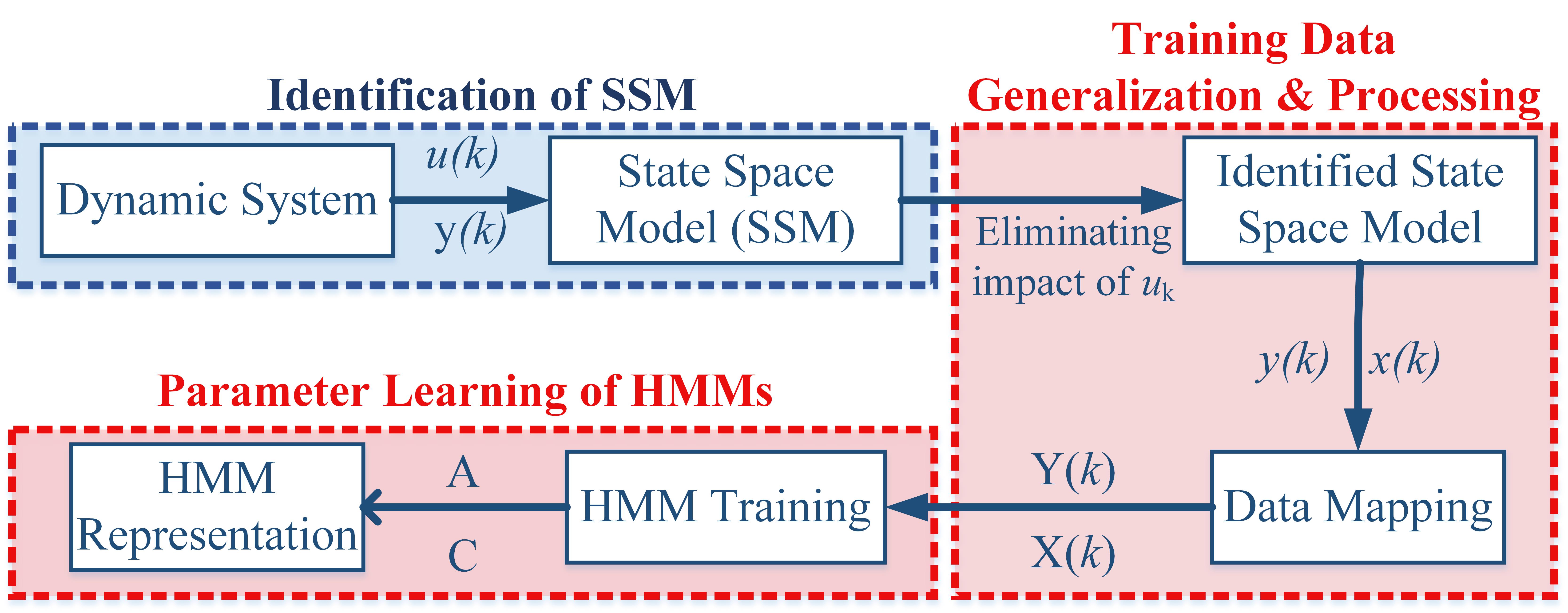}
	\caption{Schematic of the indirect method for training an HMM.}\label{fig:process}
\end{figure}

{\color{V3}Note that in equation \eqref{equ:sys_without input}, the effect of deterministic input is omitted, which can be evaluated with the knowledge of the SSM. Specifically, write
\begin{equation}\label{equ:sys1}
\begin{split}
{\rm x}(k+1) &= \mathcal{A}{\rm x}(k)+\mathcal{B}u(k)+w(k), \\
{\rm y}(k) &=\mathcal{C}{\rm x}(k)+\mathcal{D}u(k)+v(k),
\end{split}
\end{equation}
where $u_k$ denotes the deterministic input signal. For systems \eqref{equ:sys_without input} and \eqref{equ:sys1}, if the input signal $u(k)$ and the system parameters ${\color{V3}\mathcal{A}}$, ${\color{V3}\mathcal{B}}$, ${\color{V3}\mathcal{C}}$, ${\color{V3}\mathcal{D}}$ are known and the initial states of them are the same, the state and measurement $x(k), y(k)$ of system \eqref{equ:sys_without input} can be calculated based on ${{\rm x}(k),{\rm y}(k)}$ according to
	\begin{align}
	x(k) &={\rm x}(k)-\Pi(k)U(k) \label{equ:prop11}\\
	y(k) &={\rm y}(k)-{\color{V3}\mathcal{C}}\Pi(k)U(k)-{\color{V3}\mathcal{D}}u(k),\label{equ:prop12}
	\end{align}
	where
	\begin{align*}
	\Pi(k) &=\left[({\color{V3}\mathcal{A}})^{k-1}{\color{V3}\mathcal{B}}\; ({\color{V3}\mathcal{A}})^{k-2}{\color{V3}\mathcal{B}}\; {\color{V3}\ldots} \; {\color{V3}\mathcal{A}}{\color{V3}\mathcal{B}} \; {\color{V3}\mathcal{B}}\right] \end{align*}\begin{align*} 
	U(k) &=\left[u(0)\; u(1)\; {\color{V3}\ldots}\; u(k-1)\right]^\mathrm{T}.
	\end{align*}
With these relationships, Algorithm \ref{Alg:input} can be employed to calculate $x(k)$ and $y(k)$  iteratively, {\color{V1}where $D_z$ is the length of the data set}. For this reason, the effect of $u(k)$ is not considered in our work.
\begin{algorithm}[t]
	\caption{Calculation of ${x}(k)$ and ${y}(k)$ based on ${\rm x}(k)$ and ${\rm y}(k)$.}
	\begin{algorithmic}\label{Alg:input}
		\STATE$x_{\rm zs}(0)=0$;\\
		\FOR {$k={\color{V1}0}$ \TO ${\color{V1}D_z}$ }
		\STATE$x_{\rm zs}(k+1)={\color{V3}\mathcal{A}}x_{\rm zs}(k)+{\color{V3}\mathcal{B}}u(k)$;\\
		\STATE ${x}(k)={\rm x}(k)-x_{\rm zs}(k)$;\\
		\STATE ${y}(k)={\rm y}(k)-{\color{V3}\mathcal{C}}x_{\rm zs}(k)$;\\
		\ENDFOR \\
	\end{algorithmic}
\end{algorithm}

}
 
In this paper, our aim is to obtain an HMM description in the form of \eqref{equ:A} and \eqref{equ:C} for a linear Gaussian system with a  state space model in {\color{V3}\eqref{equ:sys_without input}}, with an emphasis on how to reduce the computation complexity involved in obtaining the transition matrices $A$ and $C$.

\textcolor[rgb]{0,0.267,0.588}{\section{\textsc{An Indirect Method for Learning an HMM}}}
In order to obtain the transition probability parameters of an HMM, an indirect method for training an HMM is designed in this section. 
Basically, the proposed approach is composed of three steps  (see Fig.~\ref{fig:process}). 

First, a state-space model is obtained using system identification methods (e.g., subspace methods \cite{VERHAEGENSubspace1,VERHAEGENSubspace2,VERHAEGENSubspace3}).  
Second, the obtained SSM is used as an emulator to generate the dataset $\{y(k), x(k)\}$ and $\{Y(k), X(k)\}$  needed for further parameter learning. Third, the HMM parameters $A$ and $C$ are learned based on $\{Y(k), X(k)\}$. As the aim of this work is to learn an HMM,  we assume  the first step is completed and the state space model parameters are available, and thus the remainder of this section will focus on the second and third steps.

Note that the second step is needed for two reasons: 1) the state data $x(k)$ is not available in the dataset used for system identification, and 2) the size of the dataset used for SSM identification is not large enough for HMM learning as $A$ and $C$ in \eqref{equ:A} and \eqref{equ:C} have much more parameters than those in the SSM. The dataset $\{y(k), x(k)\}$ is then processed and quantized to generate $\{Y(k), X(k)\}$, namely, the dataset used to learn HMM parameters.

\textcolor[rgb]{0,0.267,0.588}{\subsection{Generalization and Processing of Training Data}}

Comparing the HMM in \eqref{equ:A} and \eqref{equ:C} with system \eqref{equ:sys_without input}, the key observation is that both state dynamics are Markovian. However, the state of an SSM is an $n$-dimensional vector in $\mathbb{R}^n$ for an $n$th-order system while a state in an HMM is an $N$-dimensional unit vector in a finite set $S_X$. Therefore an exact mapping is impossible{\color{V3}. The} idea is to build a quantization mapping from the state and output spaces $\mathbb{R}^n$ and $\mathbb{R}^m$ of the system in \eqref{equ:sys_without input} to the state spaces $S_X$ and $S_Y$ of the HMM in \eqref{equ:A} and \eqref{equ:C}.

When ${\color{V3}\mathcal{A}}$ is stable, the probability distributions of $x_p$ converges to Gaussian distributions at the steady state
\begin{equation*}
x_p\sim N(0,(\sigma^x_p)^2)\quad\quad p\in\{1,{\color{V3}\ldots},n\}.
\end{equation*}
Now we introduce a naive quantization method to partition $\mathbb{R}$ into $N_p$ subintervals for $x_p$. We observe that $x_p$ takes values in a finite interval \cite{book2009}
\begin{equation*}
x_p\in\left[\underline x_p,\overline x_p\right]=\left[-\rho\sigma^x_p,+\rho\sigma_p^x\right],
\end{equation*}
with a large probability depending on $\rho$. This finite interval $\left[-\rho\sigma_p^x, +\rho\sigma_p^x\right]$ is then divided into $N_p-2$ equal sub-intervals$\footnote{Depending on the quantization method selected, this can be done in different ways, which is not the main focus of our work.}$, and $h_p^x$ is used to denote the length of sub-intervals. After adding $\left(-\infty, -\rho\sigma_p^x\right]$ and $\left[+\rho\sigma_p^x, +\infty\right)$ as the first and last sub-interval, respectively, we obtain $N_p$ sub-intervals. For the event that $x_p(k)$ belongs to the $\pi_p^x(k)$th sub-interval, we {\color{V2}define}
\begin{equation}\label{quanx}
X_p(k)=\delta_{N_p}^{\pi_p^x(k)}\Leftrightarrow x_p(k)\in\left[\underline
\pi_p^x(k),\overline\pi_p^x(k)\right],
\end{equation}
where $X_p(k)$ denotes the quantized $x_p(k)$, and $\underline
\pi_p^x(k)$, $\overline\pi_p^x(k)$ represent the lower and upper bounds of the sub-interval, respectively.

Similarly, $y_p$ converges to Gaussian distribution as
\begin{equation*}
y_p\sim N\left(0,(\sigma_p^y)^2\right)\quad\quad p\in\{1,{\color{V3}\ldots},m\}.
\end{equation*}
Based on the observation that 
\begin{equation*}
y_p\in\left[\underline y_p,\overline y_p\right]=\left[-\rho\sigma_p^y,+\rho\sigma_p^y\right],
\end{equation*}
the quantization of $y_p$ is {\color{V2}defined} as
\begin{equation}\label{quany}
Y_p(k)=\delta_{M_p}^{\pi_p^y(k)}\Leftrightarrow y_p(k)\in\left[\underline
\pi_p^y(k),\overline\pi_p^y(k)\right],
\end{equation}
where $Y_p(k)$ denotes the quantized $y_p(k)$, and $\underline\pi_p^y(k), \overline \pi_p^y(k)$ represent the lower and upper bound of the sub-interval, respectively. The notation $h_p^y$ is used to denote the length of sub-intervals.

Then, through combining $\left\{X_1(k),{\color{V3}\ldots},X_n(k)\right\}$, $X(k)$ can be calculated with the Kronecker product of them, which is obtained by
{\color{V2}\begin{equation}\label{kronx}
X(k)=X_1(k)\otimes X_2(k)\otimes\cdots\otimes X_n(k)=\delta_N^{\pi^x(k)}.
\end{equation}}
Similarly, we obtain
{\color{V2}\begin{equation}\label{krony}
Y(k)=Y_1(k)\otimes Y_2(k)\otimes\cdots\otimes Y_m(k)=\delta_M^{\pi^y(k)}.
\end{equation}}
The mapping between $\mathbb{R}^n, \mathbb{R}^m$ and $S_X, S_Y$ is proposed so far and Algorithm \ref{Alg:map} shows the detailed steps.
\begin{algorithm}[htp]
	\caption{Mapping algorithm of state translation from $x_k, y_k$ to $X_k, Y_k$}\label{Alg:map}
	\begin{algorithmic}[1]
		\STATE Input $x(k), y(k)$;\\
		\FOR {$p\in\{1,2,{\color{V3}\ldots},n\}$}
		\STATE $m_p^0=-\infty$;\\
		\STATE $m_p^1=\underline x_p$;\\
		\STATE $m_p^{N_p-1}=\overline {x}_p$;\\
		\STATE $m_p^{N_p}=\infty$;\\
		\FOR {$j\in\{2,3,{\color{V3}\ldots},N_p-2\}$}
		\STATE $m_p^{j}=m_p^1+(j-1)\frac{\overline x_p-\underline x_p}{N_p-2}$;\\
		\ENDFOR
		\FOR {$\pi\in\{1,2,{\color{V3}\ldots},N_p\}$}
		\IF {\color{V3}{$m_p^{\pi-1}\leq x_p(k)\leq m_p^{\pi} $}}
		\STATE  $X_p(k)=\delta_{N_p}^\pi$;
		\ENDIF
		\ENDFOR
		\ENDFOR
		\FOR {$p\in\{1,2,{\color{V3}\ldots},m\}$}
		\STATE $m_p^0=-\infty$;\\
		\STATE $m_p^1=\underline y_p$;\\
		\STATE $m_p^{M_p-1}=\overline y_p$;\\
		\STATE $m_p^{M_p}=\infty$;\\
		\FOR {$j\in\{2,3,{\color{V3}\ldots},M_p-2\}$}
		\STATE $m_p^{j}=m_p^1+(j-1)\frac{\overline y_p-\underline y_p}{M_p-2}$;\\
		\ENDFOR
		\FOR {$\pi\in\{1,2,{\color{V3}\ldots},M_p\}$}
		\IF {\color{V3}{$m_p^{\pi-1}\leq y_p(k)\leq m_p^{\pi} $}}
		\STATE  $Y_p(k)=\delta_{M_p}^\pi$;
		\ENDIF
		\ENDFOR
		\ENDFOR
		\STATE $X(k)=X_1(k)\otimes X_2(k)\otimes\cdots\otimes X_n(k)$;
		\STATE $Y(k)=Y_1(k)\otimes Y_2(k)\otimes\cdots\otimes Y_m(k)$;
		\STATE Output $X(k), Y(k)$.
	\end{algorithmic}
\end{algorithm}

{\color{V3}In Algorithm \ref{Alg:map},} $m_p^0,{\color{V3}\ldots},m_p^{N_p}$ are calculated as boundaries of sub-intervals {\color{V3}(Lines 3-9)}, based on which $x_p(k)$ is quantized to $X_p(k)$ {\color{V3}(Lines 10-14)}. Then, repeated work is performed to quantize $y_p(k)$ to $Y_p(k)$ (Lines 16-29). Finally, $X(k)$ and $Y(k)$ are obtained using equation \eqref{kronx} and \eqref{krony} (Lines 30-31). Note that the boundaries are constant for a given system, thus the boundaries can be stored to avoid repeated calculation for different data $x(k)$, $y(k)$.

This mapping approximately but uniquely associates a state of $x(k),$ $y(k)$ to a state $X(k),$ $Y(k)$, which is detailed in Algorithm \ref{Alg:map}. In this way, the training data $\left\{Y(k),X(k)\right\}$ needed to learn $A$ and $C$ are obtained based on $\left\{y(k),x(k)\right\}$.

\textcolor[rgb]{0,0.267,0.588}{\subsection{Training of an HMM: A Naive Method}}
Based on the law of large numbers, the average of the results obtained from a large number of trials should be close to the expected value \cite{book2009}. Combined with a state {\color{V3}sequence} $\{X(0), X(1), ..., X(S)\}$ and an observation {\color{V3}sequence} $\{Y(0), Y(1), ..., Y(S)\}$, the parameters ${A}$ and ${C}$ of an HMM can be evaluated as
\begin{equation}\label{equ:ber}
{a}_{i,j}^+ = \frac{X_{i,j}}{\sum\limits_{i=1}^{N}X_{i,j}},\quad
{c}_{i,j}^+ = \frac{Y_{i,j}}{\sum\limits_{i=1}^{M}Y_{i,j}},
\end{equation}
where $X_{i,j}$ represents the count of the event that $X_k$$=\delta_N^j$$\cap X_{k+1}$$=\delta_N^i$, $Y_{i,j}$ denotes the count of the event that $X_k$$=\delta_N^j$$\cap Y_{k}$$=\delta_M^i$. The notations ${a}_{i,j}^+$ and ${c}_{i,j}^+$ are used to denote the transition probability estimated according to the law of large numbers.
Then we define
\begin{equation}\label{apcp}
\begin{split}
{A}^+:=(\hat{a}_{i,j}^+)\in\mathbb{R}^{N\times N}, \quad
{C}^+:=(\hat{c}_{i,j}^+)\in\mathbb{R}^{M\times N}.
\end{split}
\end{equation}
The method of the evaluation is shown in Algorithm \ref{Alg:1}.

\begin{algorithm}[htp]
	\caption{{\color{V3}T}raining of a transition matrix ${A}^+$ and a measurement matrix ${C}^+$}\label{Alg:1}
	\begin{algorithmic}[1]
	{\color{V2}	\STATE Initialize ${A}^+, {C}^+$ and ${A}^{\rm z}, {C}^{\rm z}$;}\\
	{\color{V2}	\STATE Initialize the number of loop $L$ and the data size $D_{\rm z}$};
		\FOR {$i=1$ \TO $L$}
	{\color{V2}	\STATE Initialize ${A}^{\rm z}, {C}^{\rm z}$;}\\
	{\color{V2}	\STATE Simulate the system (\ref{equ:sys_without input}) to generate $D_{\rm z}$ new data;} (Generally, the data is different in each loop because of the disturbances.)\\
		\FOR {$k=1$ \TO $D_{\rm\color{V2} z}$}
		\STATE Translate ${\rm x}(k), {\rm y}(k)$ to ${x}(k), {y}(k)$ using Algorithm \ref{Alg:input};\\
		{\color{V2}\STATE Quantize ${x}(k), {y}(k)$ to calculate ${\pi^x(k)}$ and ${\pi^y(k)}$ according to Algorithm \ref{Alg:map};}\\
		{\color{V3}\STATE $c^{\rm z}_{\pi^y(k),\pi^x(k)}=c^{\rm z}_{\pi^y(k),\pi^x(k)}+1;$}
		\ENDFOR
		\FOR {$k=1$ \TO $D_{\rm\color{V2} z-1}$}
		{\color{V3}\STATE $a^{\rm z}_{\pi^x(k+1),\pi^x(k)}=a^{\rm z}_{\pi^x(k+1),\pi^x(k)}+1;$}
		\ENDFOR
		\STATE $A^+=A^++A^{\rm\color{V2} z}$;\\
		\STATE $C^+=C^++C^{\rm\color{V2} z}$;\\
		\ENDFOR\\
		\FOR {$j=1$ \TO $N$}\label{alg:test1}
		\IF {${\rm sum}([A^+]^j)>0$}
		\STATE $[A^+]^j=[A^+]^j\cdot({\rm sum}([A^+]^j))^{-1}$;\\
		\ENDIF
		\IF {${\rm sum}([C^+]^j)>0$}
		\STATE $[C^+]^j=[C^+]^j\cdot({\rm sum}([C^+]^j))^{-1}$;\\
		\ENDIF
		\ENDFOR\label{alg:test2}
	\end{algorithmic}
\end{algorithm}
Specifically, Algorithm \ref{Alg:1} is mainly composed of the following steps. Training parameters and transition matrices are initialized (Lines 1-2). In each loop, training data are generalized (Line 5) and processed (Lines 7-8). {\color{V2}The quantization results are used to count state transition events and stored as $A^{\rm z}, C^{\rm z}$ as sparse matrices (Line 9 and Lines 11-13). The matrices $A^{\rm z}, C^{\rm z}$} in every loop are summed up as $A^+, C^+$, respectively (Lines 14-15). Normalization of transition matrices $A^+, C^+$ is performed at the end of Algorithm \ref{Alg:1} (Lines 17-24). {\color{V3-2}The number of the parameters learned through ``exhaustive training'' in Algorithm \ref{Alg:1} equals to $\prod\limits_{p=1}^n(N_p)^2+\prod\limits_{p=1}^nN_p\prod\limits_{p=1}^mM_p$. By ``exhaustive training'', we mean that the parameter is trained/learned  based on exhaustive Monte Carlo simulation, the precision of which depends on the size of the training data, $L\times D_{\rm z}$.} 

\begin{rem}
	Several tricks are taken in Algorithm \ref{Alg:1} to reduce the time cost and errors. In view of the time cost of retrieving dense matrices, a series of sparse matrices are trained and the weighted average of them can be estimations of ${A}$ and ${C}$. The parameters are normalized at the end of Algorithm \ref{Alg:1} to reduce round-off errors.
\end{rem}

\textcolor[rgb]{0,0.267,0.588}{\subsection{Training of an HMM: An Improved Method}}
Equation (\ref{equ:ber}) and Algorithm \ref{Alg:1} provide a method to evaluate the parameters of an HMM. The evaluation will become more accurate with the increase of the data size {\color{V2}$L\times D_{\rm z}$}. Unfortunately, the time complexity of the algorithm will become unacceptable at the same time, especially when the order of the system in \eqref{equ:sys1} is high or the cardinalities $N, M$ are large. This prompts the design of learning algorithm with reduced complexity, which is discussed in this subsection.

Fortunately, we find a pattern which can be used to solve the problem. A column of the transition matrix is shown in Fig.~2 and Fig.~3, where the orders of the systems considered are 2 and 3, respectively. From these two figures, we observe that a similar pattern is shared in the curves that correspond to different blocks of the matrices.
\begin{figure}[htp]\label{fig:column of A}
	\centering
	\includegraphics[width=\linewidth]{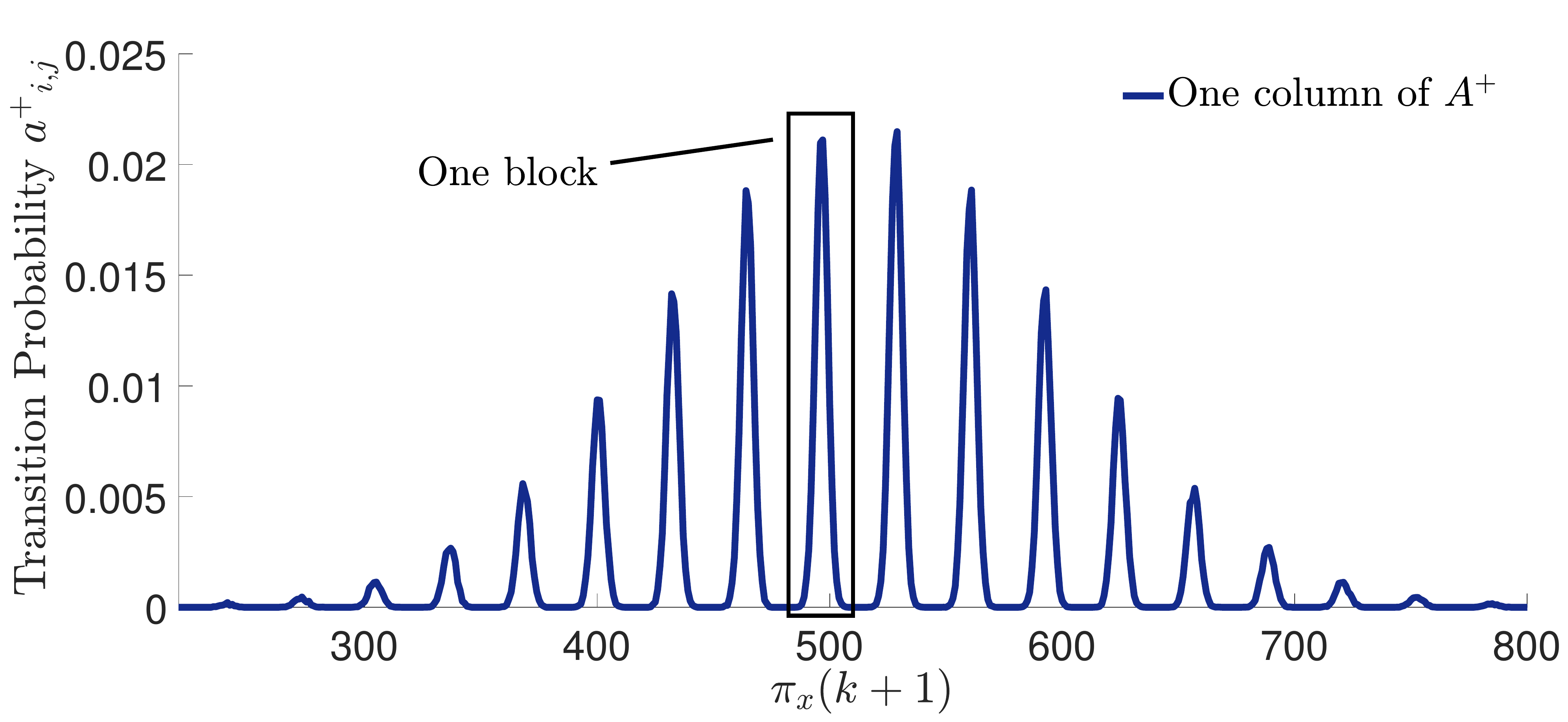}
	\caption{One column of ${\color{V1}A^+}$ for a $2$nd-order system.}
\end{figure}
\begin{figure}[htp]\label{fig:column of3}
	\centering
	\includegraphics[width=\linewidth]{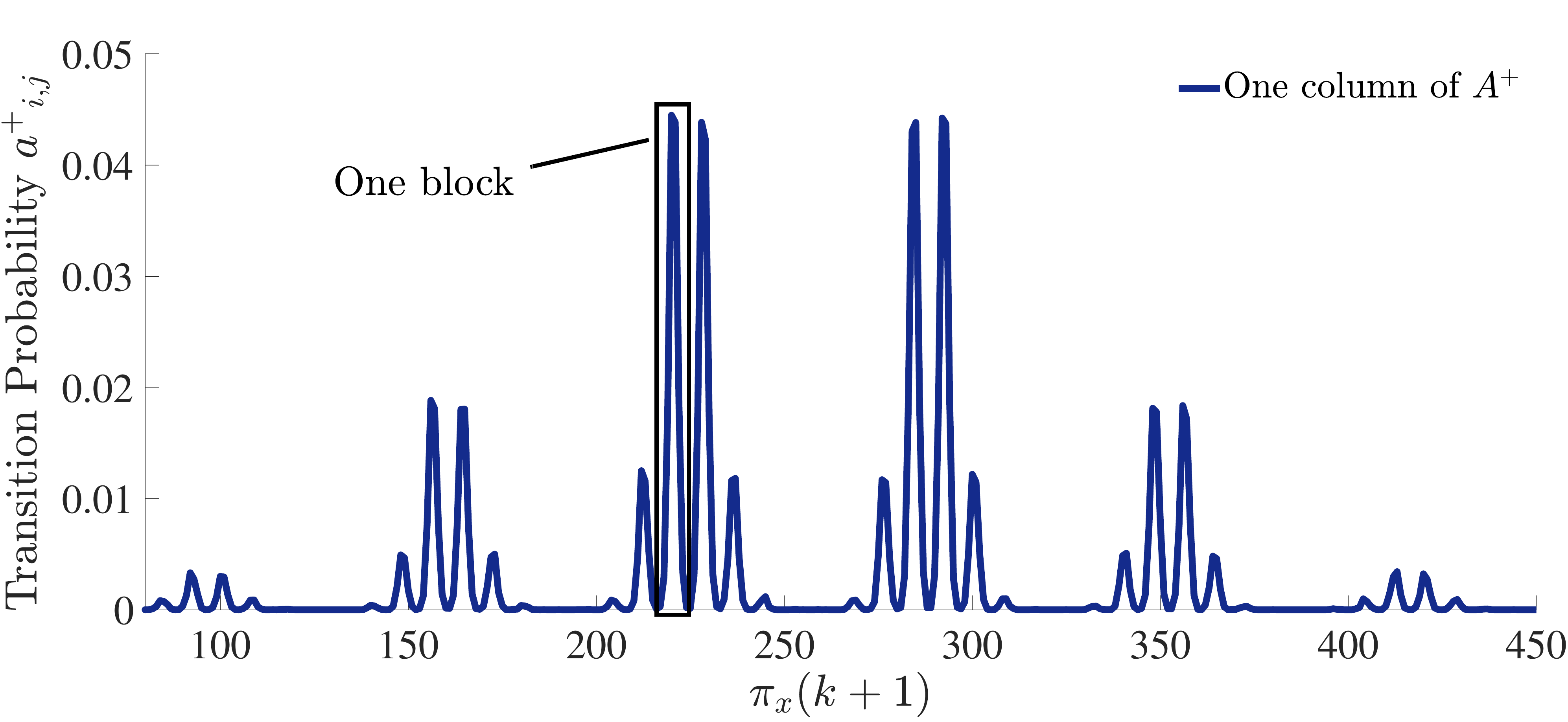}
	\caption{One column of ${\color{V1}A^+}$ for a $3$rd-order system.}
\end{figure}

{\color{black}The plots of a column of $A^+$ for different systems in Fig.~2 and Fig.~3 show that a pattern featuring a unimodal curve appears repeatedly. The periodical structural pattern arises from the mapping we constructed in the last subsection.} The results developed below will gradually reveal the cause of the pattern.

\begin{thm}\label{thm:kh}
	For an $n$th-order Gaussian system \eqref{equ:sys_without input} and an HMM determined as Algorithm 2, the transition matrix $A$ satisfies
	\begin{equation}\label{thmkhre}
	A={A}^1*{A}^2*\cdots*{A}^n,
	\end{equation}
	where
	\begin{align*}
	{a}^p_{i,j}&={\rm P}\left(X_p(k+1)=\delta_{N_p}^{i}|X(k)=\delta_N^j\right),\\
	{A}^p&=({a}^p_{i,j})\in\mathbb R^{N_p\times N}.
	\end{align*}
\end{thm}
\begin{pf}
	Considering that the covariance matrix $Q$ is a diagonal matrix, the independence between different elements leads to
	{\color{V2}\begin{align}\label{Ptime}
		&{\rm P}\left(X(k+1)=\delta_N^{\pi^x(k+1)}|X(k)=\delta_N^j\right)\\
		=&\prod\limits_{p=1}^n{\rm P}\left(X_p(k+1)=\delta_{N_p}^{\pi_p^x(k+1)}|X(k)=\delta_N^j\right),
	\end{align} }
	thus we obtain
	{\color{V2}\begin{equation}\label{equ:pf11}
	a_{\pi^x(k+1),j}=\prod\limits_{p=1}^n {a}^p_{\pi_p^x(k+1),j}.
	\end{equation}}
	Noticing that
	{\color{V2}\begin{equation*}
		X_p(k+1)=\delta_{N_p}^{\pi_p^x(k+1)},\quad\quad X(k+1)=\delta_N^{\pi^x(k+1)},
	\end{equation*}}
	the combination method \eqref{kronx} leads to
	{\color{V2}\begin{equation*}
		\delta_N^{\pi^x(k+1)}=\delta_{N_1}^{\pi_1^x(k+1)}\otimes\cdots\otimes\delta_{N_n}^{\pi_n^x(k+1)}.
	\end{equation*}}
	 Thus we obtain
	 \begin{equation*}
	 [A]^j=[A^1]^j\otimes\cdots\otimes[A^n]^j.
	 \end{equation*}
	 Then we have
	 \begin{equation*}
	 A={A}^1*{A}^2*\cdots*{A}^n
	 \end{equation*}
	 based on the definition of Khatri-Rao product, which completes the proof. \hfill\QEDclosed
\end{pf}
	{\color{black}
\begin{rem}
Note that the transition matrix $A^p, p\in\{1,{\color{V3}\ldots},n\}$ denotes the transition process from $X(k)$ to $X_p(k+1)$. The results of Theorem \ref{thm:kh} reveal the relationship between $A^p, p\in\{1,{\color{V3}\ldots},n\}$ and $A$, which indicates that the focus of parameter estimation can be turned to $A^p$ instead of $A$.
\end{rem}
}
Similarly, Theorem \ref{thm:ckh} is proposed to explore the pattern for measurement matrix $C$, {\color{V3-2}the proof of which is stated in {\color{V1}Appendix A}.}
\begin{thm}\label{thm:ckh}
	For an $n$th-order Gaussian system \eqref{equ:sys_without input} and an HMM determined as Algorithm 2, the measurement matrix $C$ satisfies
	\begin{equation*}
	C={C}^1*{C}^2*\cdots*{C}^m,
	\end{equation*}
	where
	\begin{align*}
	{c}^p_{i,j}&=P\left(Y_p(k)=\delta_{M_p}^{i}|X(k)=\delta_N^j\right),\\
	{C}^p&=({c}^p_{i,j})\in\mathbb R^{M_p\times N}.
	\end{align*}
\end{thm}

Theorems \ref{thm:kh} and \ref{thm:ckh} allows us to focus our analysis on ${A}^p,$ ${C}^p$ rather than $A, C$. {\color{V3}A} simple example of two columns of ${A}^1$ is shown in Fig.~4, which shows an obvious similarity between different columns.

\begin{figure}[htp]\label{fig:colc}
	\centering
	\includegraphics[width=\linewidth]{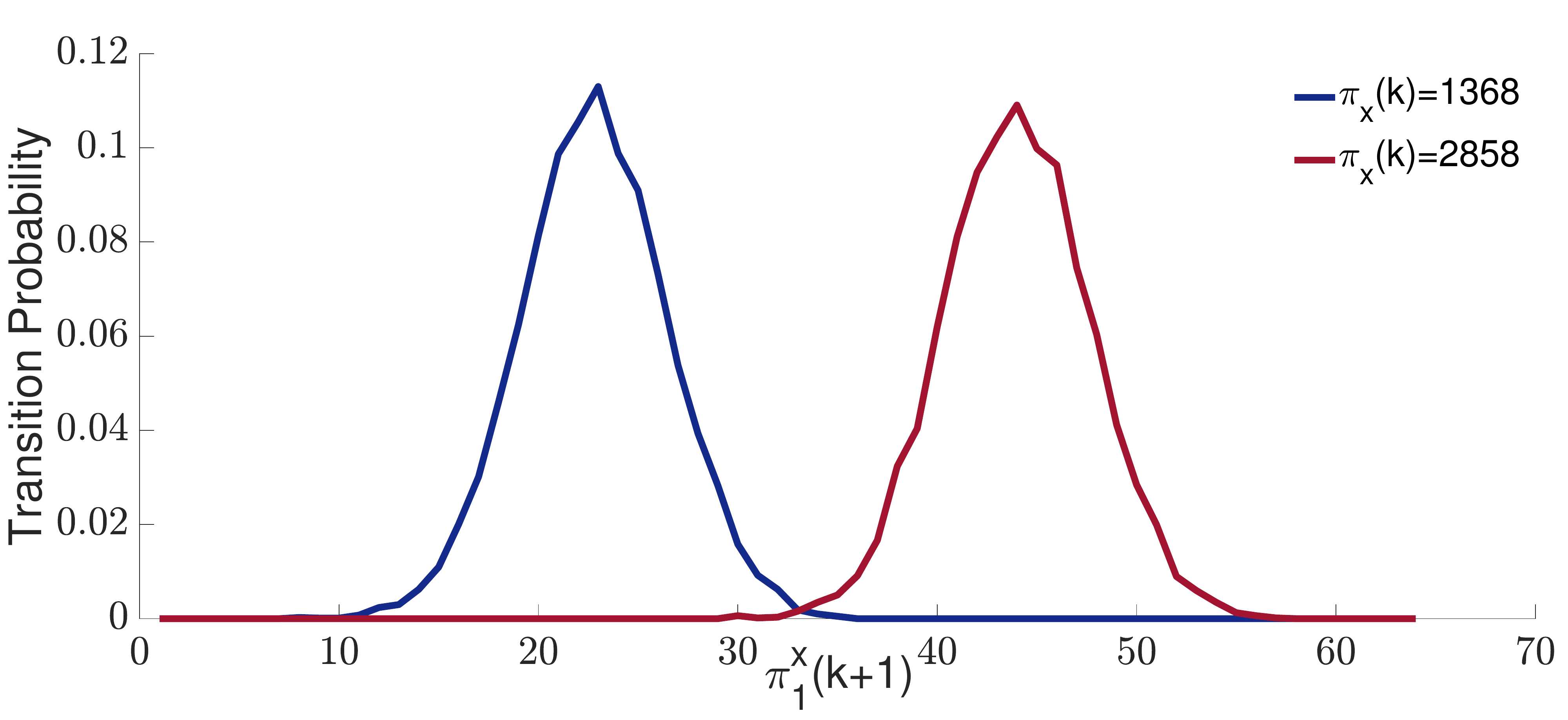}
	\caption{Two columns of ${A}^1$ for a $2$nd-order system.}
\end{figure}


The similarity between different columns of ${A}^p$ drives us to explore the relationship between different columns of transition matrix. Thus the elements of ${A}^p$ and ${C}^p$ are investigated as follows.

For a stochastic variable that corresponds to the Gaussian distribution $x\sim N(0,(\sigma)^2)$, write the probability density function as
{\color{V2}\begin{equation}
f_{\rm G}(x,\sigma):=\frac{1}{\sqrt{2\pi}\sigma}\exp(-\frac{x^2}{2\sigma^2}).
\end{equation}}
Since $x_p(k), p\in\{1,{\color{V3}\ldots},n\}$ correspond to zero-mean Gaussian distributions, the linear combinations of them also correspond to zero-mean Gaussian distributions. After writing
\begin{equation*}
\alpha^x_p(k)=[{\color{V3}\mathcal{A}}]_px(k),
\end{equation*}
the independence between $\alpha^x_p(k)$ and $w_p(k)$ leads to a two-dimensional Gaussian distribution
{\color{V2}\begin{equation}
f_p^x(\alpha^x_p(k),w_p(k)):=f_{\rm G}(\alpha^x_p(k),\bar{\sigma}_p^x)f_{\rm G}(w_p(k),q_p),
\end{equation}}
where $\bar{\sigma}_p^x$ is the variance of $\alpha^x_p(k)$.
Considering that $w_p(k)=x_p(k+1)-\alpha^x_p(k)$, we obtain
{\color{V2}\begin{align}\label{fpx}
&f_p^x(\alpha^x_p(k),w_p(k))\\
=&f_p^x(\alpha^x_p(k),x_p(k+1)-\alpha^x_p(k))\notag\\
=&f_{\rm G}(\alpha^x_p(k),\bar{\sigma}_p^x)f_{\rm G}(x_p(k+1)-\alpha^x_p(k),q_p),\notag
\end{align}}
which is a two-dimensional Gaussian distribution of $\alpha^x_p(k)$ and $x_p(k+1)$. This substitution step is necessary because the quantization method is designed based on $x(k), x(k+1)$ rather than $w(k)$.

For $X(k)=\delta_N^{j}$ and $X_p(k+1)=\delta_{N_p}^{i}$, we {\color{V2}define}
\begin{align*}
X(k)=\delta_N^{j}&\Leftrightarrow \alpha_p^x(k)\in[\underline{\alpha}_p^x(k),\overline{\alpha}_p^x(k)];\\
X_p(k+1)=\delta_{N_p}^{i}&\Leftrightarrow x_p(k+1)\in[\underline{x}_p(k+1),\overline{x}_p(k+1)],
\end{align*}
where $\underline{x}_p(k+1)$, $\overline{x}_p(k+1)$ and $\underline{\alpha}_p^x(k)$, $\overline{\alpha}_p^x(k)$ denote the boundaries of $x_p(k+1)$ and $\alpha^x_p(k)$, respectively.

 Then, the transition probability $a^p_{i,j}$ can be calculated by a integral of equation \eqref{fpx} as
\begin{align}\label{aijh}
	&{a}^p_{i,j}\\
	=&{\rm P}(X_p(k+1)=\delta_{N_p}^{i}|X(k)=\delta_N^{j})\notag\\
	=&\iint\limits_{\Omega_p^x}f_p^x(\alpha^x_p(k),x_p(k+1)-\alpha^x_p(k)){\rm d}\alpha^x_p(k){\rm d}x_p(k+1)\notag,
\end{align}
where $\Omega_p^x$ is a rectangular region defined as

\begin{align}
&\Omega_p^x\!:=\!\{\alpha_p^x(k),x_p(k\!+\!1)~|~\underline{\alpha}_p^x(k)\!\leq\!\alpha_p^x(k)\!\leq\!\overline{\alpha}_p^x(k),~~~~~~~~\nonumber\\&~~~~~~~~~~~~~~~~~~~~~~\underline{x}_p(k\!+\!1)\!\leq\! x_p(k\!+\!1)\!\leq\! \overline{x}_p(k\!+\!1)\}.
\end{align}

The results above show that, the relationship between different elements is concealed by complicated integrals. In order to avoid those complicate integrals, a method of numerical integration is adopted to analyze the transition probabilities.

Consider a continuous differentiable function $f(x,y)$ on a rectangular region $\Omega:\{\underline{x}\leq x\leq \overline{x},\underline{y}\leq y\leq \overline{y}\}$. In order to approximate $\iint\limits_\Omega f(x,y){\rm d}x{\rm d}y$ we define
{\color{V2}\begin{equation}\label{Fd}
\Psi(f(\cdot,\cdot),\underline{x},\overline{x},\underline{y},\overline{y})\!:=\!(\overline{y}-\underline{y})(\overline{x}-\underline{x})f\left(\frac{\underline{x}+\overline{x}}{2},\frac{\underline{y}+\overline{y}}{2}\right),
\end{equation}}
and the approximation error is provided in Lemma \ref{lemmati}.
\begin{lem}\label{lemmati}
The double integral of a continuous differentiable function, $\iint\limits_\Omega f(x,y){\rm d}x{\rm d}y$, can be approximated as equation \eqref{Fd} with the estimation error satisfying
{\color{V2}\begin{align*}
&\left|\iint\limits_\Omega f(x,y){\rm d}x{\rm d}y-\Psi(f(\cdot,\cdot),\underline{x},\overline{x},\underline{y},\overline{y})\right|\\
\leq&\frac{(\overline{x}-\underline{x})^3(\overline{y}-\underline{y})}{24}\max\limits_\Omega\frac{{\rm \partial^2}f(x,y)}{{\rm \partial}x^2}\\&+\frac{(\overline{x}-\underline{x})(\overline{y}-\underline{y})^3}{24}\max\limits_\Omega\frac{{\rm \partial^2}f(x,y)}{{\rm \partial}y^2}.
\end{align*}}
\end{lem}

	\begin{figure*}[b]
	\hrulefill
		\begin{align}\label{er:aijt}
	\left|\hat{a}^p_{i,j}-{a}^p_{i,j}\right|
	\leq&\frac{1}{24}(\overline{\alpha}_p^x(k)-\underline{\alpha}_p^x(k))^3(\overline{x}_p(k+1)-\underline{x}_p(k+1))\max\limits_{\Omega_p^x}\frac{{\rm \partial^2}f_p^x(\alpha_p^x(k),x_p(k+1)-\alpha_p^x(k))}{{\rm \partial}(\alpha_p^x(k))^2}\tag{28}\\&+\frac{1}{24}(\overline{\alpha}_p^x(k)-\underline{\alpha}_p^x(k))(\overline{x}_p(k+1)-\underline{x}_p(k+1))^3\max\limits_{\Omega_p^x}\frac{{\rm \partial^2}f_p^x(\alpha_p^x(k),x_p(k+1)-\alpha_p^x(k))}{{\rm \partial}(x_p(k+1))^2}.\notag
	\end{align}
\end{figure*}

\begin{pf}
{\color{black}The Taylor expansion of $f(x,y)$ about $x$ at the midpoint $\left(\frac{\overline{x}+\underline{x}}{2},y\right)$ with a Lagrange reminder term is

	\begin{align*}
	f(x,y)=&f\left(\frac{\overline{x}+\underline{x}}{2},y\right)+\frac{\partial f\left(\frac{\overline{x}+\underline{x}}{2},y\right)}{\partial x}\left(x-\frac{\overline{x}+\underline{x}}{2}\right)\\
	&\quad+\frac{1}{2}\frac{\partial^2f(\xi_1^x,y)}{\partial x^2}\left(x-\frac{\overline{x}+\underline{x}}{2}\right)^2,
	\end{align*}
where $\xi_1^x\in(\underline{x},\overline{x})$. Then, the integral of $f(x,y)$ from $\underline{x}$ to $\overline{x}$ leads to
\begin{small}
	\begin{align*}
	&\int\limits_{\underline{x}}^{\overline{x}}f(x,y){\rm d}x\!\\
	=&\int\limits_{\underline{x}}^{\overline{x}}f\left(\frac{\overline{x}+\underline{x}}{2},y\right){\rm d}x+\int\limits_{\underline{x}}^{\overline{x}}\frac{\partial f\left(\frac{\overline{x}+\underline{x}}{2},y\right)}{\partial x}\left(x-\frac{\overline{x}+\underline{x}}{2}\right){\rm d}x\\&\quad+\int\limits_{\underline{x}}^{\overline{x}}\frac{1}{2}\frac{\partial^2f(\xi_1^x,y)}{\partial x^2}\left(x-\frac{\overline{x}+\underline{x}}{2}\right)^2{\rm d}x\\
	=&(\overline{x}-\underline{x})f\left(\frac{\overline{x}+\underline{x}}{2},y\right)+\frac{\partial f\left(\frac{\overline{x}+\underline{x}}{2},y\right)}{\partial x}\int\limits_{\underline{x}}^{\overline{x}}\left(x-\frac{\overline{x}+\underline{x}}{2}\right){\rm d}x\\
	&\quad+\frac{1}{2}\frac{\partial^2f\left(\xi_1^x,y\right)}{\partial x^2}\int\limits_{\underline{x}}^{\overline{x}}\left(x-\frac{\overline{x}+\underline{x}}{2}\right)^2{\rm d}x\\
	=&(\overline{x}-\underline{x})f\left(\frac{\overline{x}+\underline{x}}{2},y\right)+0+\frac{(\overline{x}-\underline{x})^3}{24}\frac{\partial^2f(\xi_1^x,y)}{x^2}.
	\end{align*}
\end{small}
Thus we obtain
\begin{align}\label{fudx}
&\int\limits_{\underline{x}}^{\overline{x}}f(x,y){\rm d}x-(\overline{x}-\underline{x})f\left(\frac{\underline{x}+\overline{x}}{2},y\right)\\
=&\frac{(\overline{x}-\underline{x})^3}{24}\frac{\partial^2f(\xi^x_{1},y)}{\partial x^2}\notag.
\end{align}
Similarly, the Taylor expansion of $(\overline{x}-\underline{x})f\left(\frac{\overline{x}+\underline{x}}{2},y\right)$ about $y$ at the point $\left(\frac{\overline{x}+\underline{x}}{2},\frac{\overline{y}+\underline{y}}{2}\right)$ with a Lagrange reminder term is
\begin{align*}
&(\overline{x}-\underline{x})f\left(\frac{\overline{x}+\underline{x}}{2},y\right)\\
=&(\overline{x}-\underline{x})f\left(\frac{\overline{x}+\underline{x}}{2},\frac{\overline{y}+\underline{y}}{2}\right)\\
&+(\overline{x}-\underline{x})\frac{\partial f\left(\frac{\overline{x}+\underline{x}}{2},\frac{\overline{y}+\underline{y}}{2}\right)}{\partial y}\left(y-\frac{\overline{y}+\underline{y}}{2}\right)\\
&+\frac{(\overline{x}-\underline{x})}{2}\frac{\partial^2f\left(\frac{\overline{x}+\underline{x}}{2},\xi_2^y\right)}{\partial y^2}\left(y-\frac{\overline{y}-\underline{y}}{2}\right)^2,
\end{align*}
where $\xi_2^y\in\{\underline{y},\overline{y}\}$. Then, the integral of $(\overline{x}-\underline{x})f\left(\frac{\overline{x}+\underline{x}}{2},y\right)$ from $\underline{y}$ to $\overline{y}$ leads to
{\color{V2}\begin{align}\label{int2y}
&\int\limits_{\underline{y}}^{\overline{y}}(\overline{x}-\underline{x})f\left(\frac{\overline{x}+\underline{x}}{2},y\right){\rm d}y-	\Psi(f(\cdot,\cdot),\underline{x},\overline{x},\underline{y},\overline{y})\\
=&\frac{(\overline{y}-\underline{y})^3(\overline{x}-\underline{x})}{24}\frac{\partial^2f\left(\frac{\overline{x}+\underline{x}}{2},\xi^y_{2}\right)}{\partial y^2}\notag.
\end{align}}
The integral of equation \eqref{fudx} from $\underline{y}$ to $\overline{y}$ can be written as
\begin{align}\label{iintxy}
&\int\limits_{\underline{y}}^{\overline{y}}\int\limits_{\underline{x}}^{\overline{x}}f(x,y){\rm d}x{\rm d}y-\int\limits_{\underline{y}}^{\overline{y}}(\overline{x}-\underline{x})f\left(\frac{\underline{x}+\overline{x}}{2},y\right){\rm d}y\\
=&\int\limits_{\underline{y}}^{\overline{y}}\frac{(\overline{x}-\underline{x})^3}{24}\frac{\partial^2f(\xi^x_{1},y)}{\partial x^2}{\rm d}y\notag\\
=&\frac{(\overline{y}-\underline{y})(\overline{x}-\underline{x})^3}{24}\frac{\partial^2f(\xi^x_{1},\xi_1^y)}{\partial x^2}{\rm d}y\notag,
\end{align}
where $\xi_1^y\in(\underline{y},\overline{y})$.
Finally, after writing $\xi_2^x=\frac{\overline{x}+\underline{x}}{2}$ and adding equation \eqref{iintxy} to equation \eqref{int2y}, we have
{\color{V2}	\begin{align*}
		&\iint\limits_\Omega f(x,y){\rm d}x{\rm d}y-\Psi(f(\cdot,\cdot),\underline{x},\overline{x},\underline{y},\overline{y})\\
	=&\frac{(\overline{x}\!-\!\underline{x})^3(\overline{y}\!-\!\underline{y})}{24}\frac{{\rm \partial^2}f(\xi^x_{1},\xi^y_{1})}{{\rm \partial}x^2}+\frac{(\overline{x}\!-\!\underline{x})(\overline{y}\!-\!\underline{y})^3}{24}\frac{{\rm \partial^2}f(\xi^x_{2},\xi^y_{2})}{{\rm \partial}y^2}.
	\end{align*}}
	Therefore, the claim holds.} \hfill\QEDclosed
	\end{pf}

Based on Lemma \ref{lemmati}, the transition probability ${a}^p_{i,j}$ can be evaluated using numerical integral method in Lemma \ref{thm:aijt}.

\begin{lem}\label{thm:aijt}
	The transition probability ${a}^p_{i,j}$ in equation \eqref{aijh} can be approximated by $\hat{a}^p_{i,j}$ as
{\color{V2}\begin{align}\label{aijt}
		\hat{a}^p_{i,j}=\Psi(f_p^x(\cdot,\cdot),\underline{\alpha}^x_p(k),\overline{\alpha}^x_p(k),\underline{x}_p(k+1),\overline{x}_p(k+1))
	\end{align}}
	for $p\in\{1,{\color{V3}\ldots},n\}$, where
	$i=\pi_p^x(k+1), j=\pi^x(k),$ and $\pi_p^x(k),\pi_p^x(k+1)\notin\{1,N_p\},$ $p\in\{1,{\color{V3}\ldots},n\}.$ 
The estimation error satisfies equation \eqref{er:aijt}.
\end{lem}

\begin{pf}
	The claims follow easily from Lemma \ref{lemmati} by differentiation.\hfill\QEDclosed
	\end{pf}

\begin{figure*}[b]
	\hrulefill
\begin{align}\label{er:cijt}
\left|\hat{c}^p_{i,j}-{c}^p_{i,j}\right|
\leq&\frac{1}{24}(\overline{\alpha}_p^y(k)-\underline{\alpha}_p^y(k))^3(\overline{y}_p(k)-\underline{y}_p(k))\max\limits_{\Omega_p^y}\frac{{\rm \partial^2}f_p^y(\alpha_p^y(k),y_p(k)-\alpha_p^y(k))}{{\rm \partial}(\alpha_p^y(k))^2}\tag{33}\\&+\frac{1}{24}(\overline{\alpha}_p^y(k)-\underline{\alpha}_p^y(k))(\overline{y}_p(k)-\underline{y}_p(k))^3\max\limits_{\Omega_p^y}\frac{{\rm \partial^2}f_p^y(\alpha_p^y(k),y_p(k)-\alpha_p^y(k))}{{\rm \partial}(y_p(k))^2}.\notag
\end{align}
\end{figure*}

\begin{rem}\label{rem2}
	Lemma \ref{thm:aijt} proposes a simple estimation of the transition probabilities in equation \eqref{aijh}. Considering that a larger choice of $\rho$ makes the boundary conditions smaller, the boundary cases $x_p\in(-\infty,$$\underline x_{p}],$ $x_p\in[\overline x_{p},$$+\infty)$ are excluded in Lemma \ref{thm:aijt} since their influence on our estimation result can be controlled to be sufficiently small. \end{rem}

Similarly, an estimate $\hat{c}^p_{i,j}$ of transition probability ${c}^p_{i,j}$ can be formulated with a similar method. 
 {\color{V2}Write $\alpha_p^y(k)=[{\color{V3}\mathcal{C}}]_{p}x(k)$,} then the independence between $\alpha_p^y(k)$ and $v_p(k)$ leads to a two-dimensional Gaussian distribution
{\color{V2}\setcounter{equation}{28}\begin{equation}
f_p^y(\alpha_p^y(k),v_p(k)):=f_{\rm G}(\alpha_p^y(k),\bar{\sigma}_p^y)f_{\rm G}(v_p(k),r_p),
\end{equation}}
where $\bar{\sigma}_p^y$ is the variance of $\alpha_p^y(k)$.
Considering that $v_p(k)$$=y_p(k)-$$\alpha_p^y(k)$, we obtain
{\color{V2}\begin{align}\label{fpy}
&f_p^y(\alpha_p^y(k),v_p(k))\\
=&f_p^y(\alpha_p^y(k),y_p(k)-\alpha_p^y(k))\notag\\
=&f_{\rm G}(\alpha_p^y(k),\bar{\sigma}_p^y)f_{\rm G}(y_p(k)-\alpha_p^y(k),r_p),\notag
\end{align}}
which is a two-dimensional Gaussian distribution of $\alpha_p^y(k)$ and $y_p(k)$.

For $X(k)=\delta_N^{j}$ and $Y_p(k)=\delta_{M_p}^{i}$, we {\color{V2}define}
	\begin{align*}
	X(k)=\delta_N^{j}&\Leftrightarrow \alpha_p^y(k)\in[\underline{\alpha}_p^y(k),\overline{\alpha}_p^y(k)];\\
	Y_p(k)=\delta_{M_p}^{i}&\Leftrightarrow y_p(k)\in[\underline{y}_p(k),\overline{y}_p(k)],
	\end{align*}
where $\underline{y}_p(k)$, $\overline{y}_p(k)$ and $\underline{\alpha}_p^y(k)$, $\overline{\alpha}_p^y(k)$ denote the boundaries of $y_p(k)$ and $\alpha^y_p(k)$, respectively.
	
	Then, the transition probability $c^p_{i,j}$ can be calculated by a integral of equation \eqref{fpy} as
\begin{align}\label{cijh}
&{c}^p_{i,j}\\
=&{\rm P}(Y_p(k)=\delta_{M_p}^{i}|X(k)=\delta_N^{j})\notag\\
=&\iint\limits_{\Omega_p^y}f_p^y(\alpha_p^y(k),v_p(k)){\rm d}\alpha_p^y(k){\rm d}y_p(k)\notag,
\end{align}
	where $\Omega_p^y$ is a rectangular region as
	{\small
		\begin{equation*}
		\Omega_p^y\!:\!\{\underline{\alpha}_p^y(k)\!\leq\!\alpha_p^y(k)\!\leq\!\overline{\alpha}_p^y(k),\underline{y}_p(k)\!\leq\! y_p(k)\!\leq\! \overline{y}_p(k)\}.
		\end{equation*}}

 Based on Lemma \ref{lemmati}, ${c}^p_{i,j}$ can be evaluated using numerical integral method as follows.

\begin{lem}\label{thm:cijt}
	The transition probability ${c}^p_{i,j}$ shown in equation \eqref{cijh} can be approximated by $\hat{c}^p_{i,j}$ as
{\color{V2}	\begin{align}\label{cijt}
	\hat{c}^p_{i,j}
	=\Psi(f_p^y(\cdot,\cdot), \underline{\alpha}_p^y(k),\overline{\alpha}_p^y(k),\underline{y}_p(k),\overline{y}_p(k))
	\end{align}}
for $p\in\{1,{\color{V3}\ldots},m\}$, where
	$i=\pi_p^y(k), j=\pi^x(k)$, and $\pi_p^x(k) \notin\{1,N_p\},\quad p\in\{1,{\color{V3}\ldots},n\};$
	$\pi_p^y(k) \notin\{1,M_p\},$ $\quad p\in\{1,{\color{V3}\ldots},m\}.$ The estimation error satisfies equation \eqref{er:cijt}.	
\end{lem}
\begin{pf}
	The claims follow easily from Lemma \ref{lemmati} by differentiation.\hfill\QEDclosed
\end{pf}

\begin{rem}
	Lemma \ref{thm:cijt} proposes a simple estimation of the transition probabilities in equation \eqref{cijh}. Similar to the analysis in Remark~\ref{rem2}, 
	the boundary cases $x_p\in(-\infty,$$\underline x_{p}],$ $x_p\in[\overline x_{p},$$+\infty)$, $y_p\in(-\infty,$$\underline y_{p}],$ $y_p\in[\overline y_{p},$$+\infty)$  are excluded as their influence on estimation can be sufficiently small. \end{rem}

Lemmas \ref{thm:aijt} and \ref{thm:cijt} provide numerical integral representation of the transition probabilities, which can be used to investigate the relationship between different columns of ${A}^p$ or ${C}^p$. {\color{V2}Note that variables for standard columns are marked by `${\rm\textsc s}$' and variables for target columns are marked by `${\rm \textsc t}$', where a standard column is a column which is well trained and a target column is a column needs to be estimated.}
{\color{V2}\begin{thm}\label{colaij}
	For a standard column $[A^p]^{j_{\rm\textsc s}}$ and a target column $[A^p]^{j_{\rm\textsc t}}$, the elements of the two columns satisfy
\setcounter{equation}{33}	\begin{equation}
	\lim\limits_{h_p^x\rightarrow 0}(\hat{a}^p_{i_{\rm\textsc t},j_{\rm\textsc t}}-\kappa_x\hat{a}^p_{i_{\rm\textsc s}-g_p^x,j_{\rm\textsc s}})=0,\label{ata}
	\end{equation}
	where $i_{\rm\textsc t}=i_{\rm\textsc s}$, and $\kappa_x$ is a constant as 
	{\color{V2}\begin{footnotesize}
	\begin{equation}\label{kappax}
	\kappa_x\!\!=\!\!\frac{\left(\overline{\alpha}_p^x(k;{\rm\textsc t})\!-\!\underline{\alpha}_p^x(k;{\rm\textsc t})\!\right)\!\!(\overline{x}_p(k\!+\!1;{\rm\textsc t})\!-\!\underline{x}_p(k\!+\!1;{\rm\textsc t}))f_{\rm G}\left(\mu_{\rm\textsc t}^x,\bar{\sigma}_k^p\right)}{\left(\overline{\alpha}_p^x(k;{\rm\textsc s})\!-\!\underline{\alpha}_p^x(k;{\rm\textsc s})\!\right)\!\!(\overline{x}_p(k\!+\!1;{\rm\textsc s})\!-\!\underline{x}_p(k\!+\!1;{\rm\textsc s}))f_{\rm G}\left(\mu_{\rm\textsc s}^x,\bar{\sigma}_k^p\right)}.
	\end{equation}
	\end{footnotesize}}
 The parameters $j_{\rm\textsc t}$, $j_{\rm\textsc s}$ and $g_p^x$ in equation \eqref{ata} satisfy
 \begin{align*}
 \delta_N^{j_{\rm\textsc t}} &= \delta_{N_1}^{\pi_{1}^x(k;{\rm\textsc t})}\otimes\cdots\otimes\delta_{N_n}^{\pi_n^x(k;{\rm\textsc t})} ,\\
 \delta_N^{j_{\rm\textsc s}} &= \delta_{N_1}^{\pi_1^x(k;{\rm\textsc s})}\otimes\cdots\otimes\delta_{N_n}^{\pi_n^x(k;{\rm\textsc s})} ,\\
 g_p^x&= \left\lfloor\frac{\mu_{\rm\textsc t}^x-\mu_{\rm\textsc s}^x}{h_p^x} \right\rfloor,\\
\mu_{\rm\textsc t}^x &=\frac{1}{2}\left[\overline{\alpha}_p^x(k;{\rm\textsc t})+\underline{\alpha}_p^x(k;{\rm\textsc t})\right],\\
\mu_{\rm\textsc s}^x &=\frac{1}{2}\left[\overline{\alpha}_p^x(k;{\rm\textsc s})+\underline{\alpha}_p^x(k;{\rm\textsc s})\right],
 \end{align*}
 with $\overline{\pi}_p^x(k+1;{\rm\textsc t})$$,\overline{\pi}_p^x(k+1;{\rm\textsc s})$$,\underline{\pi}_p^x(k+1;{\rm\textsc t})$$,\underline{\pi}_p^x(k+1;{\rm\textsc s})$$\notin\{+\infty,-\infty\}$, and $\overline{\pi}_p^x(k;{\rm\textsc t}),\overline{\pi}_p^x(k;{\rm\textsc s}),\underline{\pi}_p^x(k;{\rm\textsc t}),\underline{\pi}_p^x(k;{\rm\textsc s})\notin\{+\infty,-\infty\}$.
\end{thm}
\begin{pf}
	Based on Lemma \ref{thm:aijt}, we obtain
	\begin{footnotesize}
	\begin{align}	
	\hat{a}^p_{i_{\rm\textsc t},j_{\rm\textsc t}}=&\left(\overline{\alpha}_p^x(k;{\rm\textsc t})-\underline{\alpha}_p^x(k;{\rm\textsc t})\right)(\overline{x}_p(k+1;{\rm\textsc t})-\underline{x}_p(k+1;{\rm\textsc t}))\notag\\
		\quad&f_p^x\left(\mu_{\rm\textsc t}^x,\frac{\overline{x}_p(k+1;{\rm\textsc t})+\underline{x}_p(k+1;{\rm\textsc t})}{2}-\mu_{\rm\textsc t}^x\right);\label{fa1}\\
	\hat{a}^p_{i_{\rm\textsc s},j_{\rm\textsc s}}=&\left(\overline{\alpha}_p^x(k;{\rm\textsc s})-\underline{\alpha}_p^x(k;{\rm\textsc s})\right)(\overline{x}_p(k+1;{\rm\textsc s})-\underline{x}_p(k+1;{\rm\textsc s}))\notag\\
	\quad&f_p^x\left(\mu_{\rm\textsc s}^x,\frac{\overline{x}_p(k+1;{\rm\textsc s})+\underline{x}_p(k+1;{\rm\textsc s})}{2}-\mu_{\rm\textsc s}^x\right).\label{fa2}	
	\end{align}
	\end{footnotesize}
According to equation  \eqref{fpx}, we have
\begin{small}
\begin{align}
&f_p^x\left(\mu_{\rm\textsc t}^x,\frac{\overline{x}_p(k+1;{\rm\textsc t})+\underline{x}_p(k+1;{\rm\textsc t})}{2}-\mu_{\rm\textsc t}^x\right)\notag\\
=&
f_{\rm G}\left(\mu_{\rm\textsc t}^x,\bar{\sigma}_p^x\right)
\cdot f_{\rm G}\left(\frac{\overline{x}_p(k+1;{\rm\textsc t})+\underline{x}_p(k+1;{\rm\textsc t})}{2}-\mu_{\rm\textsc t}^x,q_p\!\right);\label{fa3}\\
&f_p^x\left(\mu_{\rm\textsc s}^x,\frac{\overline{x}_p(k+1;{\rm\textsc s})+\underline{x}_p(k+1;{\rm\textsc s})}{2}-\mu_{\rm\textsc s}^x\right)\notag\\
=&
f_{\rm G}\left(\mu_{\rm\textsc s}^x,\bar{\sigma}_p^x\right)
\cdot f_{\rm G}\left(\frac{\overline{x}_p(k\!+\!1;{\rm\textsc s})\!+\!\underline{x}_p(k\!+\!1;{\rm\textsc s})}{2}\!-\!\mu_{\rm\textsc s}^x,q_p\!\right).\label{fa4}
\end{align}
\end{small}

Combining equations \eqref{fa1} and \eqref{fa3}, \eqref{fa2} and \eqref{fa4}, we write
\begin{small}
\begin{align}
\hat{a}^p_{i_{\rm\textsc t},j_{\rm\textsc t}}&\!=\!\tau_p^x(k;{\rm\textsc t})f_{\rm G}\left(\frac{\overline{x}_p(k\!+\!1;{\rm\textsc t})\!+\!\underline{x}_p(k\!+\!1;{\rm\textsc t})}{2}\!-\!\mu_{\rm\textsc t}^x,q_p\!\!\right);\label{apt}\\
\hat{a}^p_{i_{\rm\textsc s},j_{\rm\textsc s}}&\!=\!\tau_p^x(k;\!{\rm\textsc s})f_{\rm G}\left(\frac{\overline{x}_p(k\!+\!1;{\rm\textsc s})\!+\!\underline{x}_p(k\!+\!1;{\rm\textsc s})}{2}\!-\!\mu_{\rm\textsc s}^x,q_p\!\!\right)\!,\label{aps}
\end{align}
\end{small}
where $\tau_p^x(k;{\rm\textsc t})$ and $\tau_p^x(k;{\rm\textsc s})$ are constants as
{\small
\begin{align*}
\tau_p^x(k;{\rm\textsc t})&\!=\!\left(\!\overline{\alpha}_p^x(k;\!{\rm\textsc t})\!-\!\underline{\alpha}_p^x(k;\!{\rm\textsc t})\!\right)\!(\overline{x}_p(k\!+\!1;\!{\rm\textsc t})\!-\!\underline{x}_p(k\!+\!1;\!{\rm\textsc t})\!)f_{\rm G}\left(\mu_{\rm\textsc t}^x,\!\bar{\sigma}_p^x\right)\\
\tau_p^x(k;{\rm\textsc s})&\!=\!\left(\!\overline{\alpha}_p^x(k;\!{\rm\textsc s})\!-\!\underline{\alpha}_p^x(k;\!{\rm\textsc s})\!\right)\!\!(\overline{x}_p(k\!+\!\!1;\!{\rm\textsc s})\!-\!\underline{x}_p(k\!+\!\!1;\!{\rm\textsc s})\!)f_{\rm G}\!\left(\mu_{\rm\textsc s}^x,\!\bar{\sigma}_p^x\right).
\end{align*}}
Equations \eqref{apt} and \eqref{aps} show that, a transition probability of the target column $\hat{a}^p_{i_{\rm\textsc t},j_{\rm\textsc t}}$ can be seen as a product of a constant $\tau_p^x(k;{\rm\textsc t})$ (called constant part) and a Gaussian distribution function with variance $q_p^2$ (called Gaussian part), while a transition probability of the standard column $\hat{a}^p_{i_{\rm\textsc s},j_{\rm\textsc s}}$ can be seen as a product of a constant $\tau_p^x(k;{\rm\textsc s})$ (constant part) and a Gaussian distribution function with variance $q_p^2$ (Gaussian part). Thus a constant $\kappa_x$ is formed as $\kappa_x=\frac{\tau_p^x(k;{\rm\textsc t})}{\tau_p^x(k;{\rm\textsc s})}$ to close the gap between the constant parts of different columns.

Then, for Gaussian parts of $\hat{a}^p_{i_{\rm\textsc s},j_{\rm\textsc s}}$ and $\hat{a}^p_{i_{\rm\textsc t},j_{\rm\textsc t}}$, consider the elements of the standard column and the target column in a same row, where $i_{\rm\textsc t}=i_{\rm\textsc s}$, $\overline{x}_p(k+1;{\rm\textsc t})=\overline{x}_p(k+1;{\rm\textsc s})$, $\underline{x}_p(k+1;{\rm\textsc t})=\underline{x}_p(k+1;{\rm\textsc s})$. It is obvious that
\begin{align}\label{utux}
&f_{\rm G}\left(\frac{\overline{x}_p(k+1;{\rm \textsc t})+\underline{x}_p(k+1;{\rm \textsc t})}{2}\!-\!\mu_{\rm \textsc t}^x,q_p\!\right)\\
=&f_{\rm G}\left(\frac{\overline{x}_p(k+1;{\rm \textsc s})+\underline{x}_p(k+1;{\rm\textsc s})}{2}\!-\!\mu_{\rm \textsc s}^x-(\mu_{\rm \textsc t}^x-\mu_{\rm\textsc s}^x),q_p\!\right).\notag
\end{align}
For quantized transition probabilities, $(\mu_{\rm\textsc t}^x-\mu_{\rm\textsc s}^x)$ needs to be quantized to a integer multiples of $h_p^x$, where $h_p^x=\overline{x}_p(k)-\underline{x}_p(k)$ is the length of quantization sub-intervals in \eqref{quanx}. Therefore, $(\mu_{\rm\textsc t}^x-\mu_{\rm\textsc s}^x)$ is quantized as $g_p^xh_p^x$, where $g_p^x=\left\lfloor\frac{\mu_{\rm\textsc t}^x-\mu_{\rm\textsc s}^x}{h_p^x} \right\rfloor$.
Since
\begin{align*}
\left|(\mu_{\rm\textsc t}^x-\mu_{\rm\textsc s}^x)-g_p^xh_p^x\right|
=\left|(\mu_{\rm\textsc t}^x-\mu_{\rm\textsc s}^x)-\left\lfloor\frac{\mu_{\rm\textsc t}^x-\mu_{\rm\textsc s}^x}{h_p^x} \right\rfloor h_p^x\right|
\leq h_p^x,
\end{align*}
we obtain
\begin{align*}
&\lim\limits_{h_p^x\rightarrow 0}|(\mu_{\rm\textsc t}^x-\mu_{\rm\textsc s}^x)-g_p^xh_p^x|=0,\\
&\lim\limits_{h_p^x\rightarrow 0}\left|\left[\frac{\overline{x}_p(k+1;{\rm\textsc s})\!+\!\underline{x}_p(k+1;{\rm\textsc s})}{2}\!-\!\mu_{\rm\textsc s}^x-(\mu_{\rm\textsc t}^x-\mu_{\rm\textsc s}^x)\right]\right.\\
&\quad-\left.\left[\frac{\overline{x}_p(k+1;{\rm\textsc s})+\underline{x}_p(k+1;{\rm\textsc s})}{2}\!-\!\mu_{\rm\textsc s}^x-g_p^xh_p^x\right]\right|=0.
\end{align*}
Then the uniform continuity of the Gaussian distribution function leads to
\begin{small}
\begin{align}
&\lim\limits_{h_p^x\rightarrow 0}\left|f_{\rm G}\!\left(\frac{\overline{x}_p(k+1;{\rm\textsc s})\!+\!\underline{x}_p(k+1;{\rm\textsc s})}{2}\!-\!\mu_{\rm\textsc s}^x-(\mu_{\rm\textsc t}^x-\mu_{\rm\textsc s}^x),q_p\!\right)\right.\notag\\
&\;\left. -f_{\rm G}\!\left(\frac{\overline{x}_p(k+1;{\rm\textsc s})+\underline{x}_p(k+1;{\rm\textsc s})}{2}\!-\!\mu_{\rm\textsc s}^x-g_p^xh_p^x,q_p\right)\right|=0.\label{lim}
\end{align}
\end{small}
Note that
\begin{footnotesize}
\begin{align}
&\hat{a}^p_{i_{\rm\textsc t},j_{\rm\textsc t}}\label{aptk}\\
=&\tau_p^x(k;{\rm\textsc t})f_{\rm G}\left(\frac{\overline{x}_p(k\!+\!1;{\rm\textsc t})\!+\!\underline{x}_p(k\!+\!1;{\rm\textsc t})}{2}\!-\!\mu_{\rm\textsc t}^x,q_p\!\!\right);\notag\\
&\kappa_x\hat{a}^p_{i_{\rm\textsc s}-g_p^x,j_{\rm\textsc s}}\label{tapsk}\\
=&\frac{\tau_p^x(k;{\rm\textsc t})}{\tau_p^x(k;{\rm\textsc s})}\tau_p^x(k;\!{\rm\textsc s})f_{\rm G}\left(\frac{\overline{x}_p(k\!+\!1;{\rm\textsc s})\!+\!\underline{x}_p(k\!+\!1;{\rm\textsc s})}{2}\!-\!\mu_{\rm\textsc s}^x\!-\!g_p^xh_p^x,q_p\!\!\right).\notag
\end{align}
\end{footnotesize}
Equation \eqref{ata} can be obtained combining equations \eqref{utux}-\eqref{tapsk}. Thus the claim holds.	\hfill\QEDclosed

\end{pf}

Similar to Theorem \ref{colaij}, Theorem \ref{colcij} is developed for $\hat{c}^p_{i_{\rm\textsc t},j_{\rm\textsc t}}$, which is proved in {\color{V1}Appendix B}.
\begin{thm}\label{colcij}
	For a standard column $[C^p]^{j_{\rm\textsc s}}$ and a target column $[C^p]^{j_{\rm\textsc t}}$, the elements of two columns satisfy
	\begin{equation}\label{limyth}
	\lim\limits_{h_p^y\rightarrow 0}(\hat{c}^p_{i_{\rm\textsc t},j_{\rm\textsc t}}-\kappa_y\hat{c}^p_{i_{\rm\textsc s}-g_p^y,j_{\rm\textsc s}})=0,
	\end{equation}
	where $i_{\rm\textsc t}=i_{\rm\textsc s}$, and $\kappa_y$ is a constant as
	\begin{small}
	\begin{equation}\label{kappay}
	\kappa_y\!=\!\frac{\left(\overline{\alpha}_p^y(k;{\rm\textsc t})\!-\!\underline{\alpha}_p^y(k;{\rm\textsc t})\!\right)\!\!(\overline{y}_p(k;{\rm\textsc t})\!-\!\underline{y}_p(k;{\rm\textsc t}))f_{\rm G}\left(\mu_{\rm\textsc t}^y,\bar{\sigma}_p^y\right)}{\left(\overline{\alpha}_p^y(k;{\rm\textsc s})\!-\!\underline{\alpha}_p^y(k;{\rm\textsc s})\!\right)\!\!(\overline{y}_p(k;{\rm\textsc s})\!-\!\underline{y}_p(k;{\rm\textsc s}))f_{\rm G}\left(\mu_{\rm\textsc s}^y,\bar{\sigma}_p^y\right)}.
	\end{equation}
	\end{small}
	The parameters $j_{\rm\textsc t}$, $j_{\rm\textsc s}$ and $g_p^y$ in \eqref{limyth} satisfy
	\begin{align*}
	\delta_N^{j_{\rm\textsc t}} &= \delta_{N_1}^{\pi_1^x(k;{\rm\textsc t})}\otimes\cdots\otimes\delta_{N_n}^{\pi_n^x(k;{\rm\textsc t})} ,\\
	\delta_N^{j_{\rm\textsc s}} &= \delta_{N_1}^{\pi_1^x(k;{\rm\textsc s})}\otimes\cdots\otimes\delta_{N_n}^{\pi_n^x(k;{\rm\textsc s})} ,\\
	g_p^y &= \left\lfloor\frac{\mu_{\rm\textsc t}^y-\mu_{\rm\textsc s}^y}{h_p^y} \right\rfloor ,\\
	\mu_{\rm\textsc t}^y &=\frac{1}{2}\left[\overline{\alpha}_p^y(k;{\rm\textsc t})+\underline{\alpha}_p^y(k;{\rm\textsc t})\right],\\
	\mu_{\rm\textsc s}^y &=\frac{1}{2}\left[\overline{\alpha}_p^y(k;{\rm\textsc s})+\underline{\alpha}_p^y(k;{\rm\textsc s})\right],
	\end{align*}
	with $\overline{\pi}_p^y(k;{\rm\textsc t})$$,\overline{\pi}_p^y(k;{\rm\textsc s})$$,\underline{\pi}_p^y(k;{\rm\textsc t})$$,\underline{\pi}_p^y(k;{\rm\textsc s})$$\;\notin\{+\infty,-\infty\}$, and $\overline{\pi}_p^x(k;{\rm\textsc t}),\overline{\pi}_p^x(k;{\rm\textsc s}),\underline{\pi}_p^x(k;{\rm\textsc t}),\underline{\pi}_p^x(k;{\rm\textsc s})\notin\{+\infty,-\infty\}$.
\end{thm}}

Based on {\color{V3}the} theorems and lemmas above, an improved algorithm is designed as Algorithm \ref{Alg:kha}, which provides us an improved method that utilizes the periodic structural characteristics adequately. {\color{black}Since only one column of each small matrices ${A}^1,{\color{V3}\ldots},{A}^n, {C}^1,{\color{V3}\ldots},{C}^m$ mentioned in Theorem \ref{thm:kh} needs to be trained in this case,   the number of parameters need to be evaluated is significantly reduced for Algorithm \ref{Alg:kha}.}
\begin{algorithm}[htp]
	\caption{Reduced-complexity training of a transition matrix ${A}^-$ and a measurement matrix ${C}^-$}\label{Alg:kha}
	\begin{algorithmic}[1]
		\STATE Initialize ${A}^{-}, {C}^{-}$;\\
		{\color{V2}\STATE Initialize the number of loop $L$ and fix the data size $D_{\rm z}=2$;}\\
		\FOR {$i=1$ \TO $L$}
		\STATE Simulate the system (\ref{equ:sys_without input}) to generate $D_{\rm\color{V2} z}$ new data with initial state equals to zero vector; (Generally, the data is different in each loop because of the disturbances.)\\
		\FOR {$k=1$ \TO $D_{\rm\color{V2} z}$}
		\STATE Translate ${\rm x}(k), {\rm y}(k)$ to ${x}(k), {y}(k)$ using Algorithm \ref{Alg:input};\\
		\FOR {$p=1$ \TO $n$, $q=1$ \TO $m$}
		\STATE Quantize ${x}_p(k), {y_q}(k)$ to calculate $\pi_p^x(k)$ and $\pi_q^y(k)$ according to Algorithm \ref{Alg:map};\\
		\ENDFOR
		{\color{V2}\STATE Calculate $\pi^x(k)$ according to Theorem \ref{thm:kh};}
		\ENDFOR
		\STATE $k=1$;
		{\color{V3}\FOR {$p=1$ \TO $m$}
	{\color{V3}	\STATE ${c}^p_{\pi_p^y(k),\pi^{x}(k)}={c}^p_{\pi_p^y(k),\pi^{x}(k)}+1$};\\
		\ENDFOR
	\FOR {$p=1$ \TO $n$}
	{\color{V3}\STATE ${a}^p_{\pi_p^x(k+1),\pi^{x}(k)}={a}^p_{\pi_p^x(k+1),\pi^{x}(k)}+1$};\\
	\ENDFOR}
		\ENDFOR
		{\color{V3-2}\FOR {$j\in\{1,\ldots,\pi^{x}(k)-1,\pi^{x}(k)+1,N\}$}
		\FOR {$p=1$ \TO $n$}	
		\FOR {$i=1$ \TO $N_p$}
		\IF {$1\leq i-g_p^x\leq N_p$}
		\STATE$a^p_{i,j}=a^p_{i-g_p^x,\pi^{x}(k)}$;
		\ENDIF
		\ENDFOR
		\ENDFOR
		\FOR {$p=1$ \TO $m$}	
		\FOR {$i=1$ \TO $ M_p$}
		\IF{$1\leq i-g_p^y\leq M_p$}
		\STATE$c^p_{i,j}=c^p_{i-g_p^y,\pi^{x}(k)}$;
		\ENDIF
		\ENDFOR
		\ENDFOR
		\ENDFOR}
		{\color{V2}\STATE $A^-={A}^1*\cdots*{A}^n$;}\\
		{\color{V2}\STATE $C^-={C}^1*\cdots*{C}^m$;}\\
		\STATE Normalize ${A}^{-}$ and ${C}^{-}$ as lines \ref{alg:test1} to \ref{alg:test2} in Algorithm \ref{Alg:1};
	\end{algorithmic}
\end{algorithm}

Specifically, Algorithm \ref{Alg:kha} is mainly composed of the following
steps. {\color{V3}First}, training data are generalized and processed based on Algorithms \ref{Alg:input} and \ref{Alg:map} (Lines 4-11). {\color{V3}Second}, the state transition process is counted and recorded in the $\pi^x(1)$th columns of matrices {\color{V3-2}$A^p$}, $p\in\{1,{\color{V3}\ldots},n\}$ and {\color{V3-2}$C^p$}, $p\in\{1,{\color{V3}\ldots},m\}$ as standard columns {\color{V3}(Lines 12-18)}. {\color{V3}Third}, other columns of transition matrices are calculated based on the standard columns using Theorems \ref{colaij} and \ref{colcij} {\color{V3-2}(Lines 20-35)}. Then, the transition matrix $A^-$ and measurement matrix $C^-$ are obtained by the Khatri-Rao product of {\color{V3-2}$A^p$}, $p\in\{1,{\color{V3}\ldots},n\}$ and {\color{V3-2}$C^p$}, $p\in\{1,{\color{V3}\ldots},m\}$, respectively {\color{V3-2}(Lines 36-37)}. The theoretical basis is provided by Theorems \ref{thm:kh} and \ref{thm:ckh}. Finally, the transition matrices $A^-$ and $C^-$ are normalized {\color{V3-2}(Line 38)}.

{\color{V3-2}Noticeably, from Lines 12-18 of Algorithm \ref{Alg:kha}, only the $\pi^x(1)$th column of matrices $A^1,\ldots,A^n,C^1,\ldots,C^m$ are learned through exhaustive training, based on which the entire transition matrices $A, C$ are further calculated. The number of the parameters learned through exhaustive training in Algorithm \ref{Alg:kha} equals to $\sum\limits_{p=1}^nN_p+\sum\limits_{p=1}^mM_p$, which is much smaller than Algorithm \ref{Alg:1}. As a consequence, the size of training data $L\times D_{\rm z}$ needed in Algorithm \ref{Alg:kha} is also much smaller than that of Algorithm \ref{Alg:1}.}

\begin{rem}
Since the normalization of the matrices is performed at the end of Algorithm \ref{Alg:kha}, multiplying constants for every element of a column before normalization would not change the result. Therefore,   $\kappa_x$ and $\kappa_y$ can take constant values, e.g.,  $\kappa_x=\kappa_y=1$.
\end{rem}

\textcolor[rgb]{0,0.267,0.588}{\section{\textsc{Numerical Analysis}}}\label{applications}
In this section, the HMM learning approach proposed in this work is applied to event-based state estimation with an unreliable communication channel through numerical analysis.
{\color{black}First, the performance of the low-complexity learning algorithm (Algorithm~\ref{Alg:kha}) is compared with that  of Algorithm \ref{Alg:1} to verify the validity of parameter estimation methods developed. Then, the HMMs learned using the proposed algorithms are applied to event-triggered state estimation and the estimation results are compared with existing results in the literature. For comparison purpose, we ignore the numerical errors caused by system identification and assume the model obtained in (\ref{equ:sys_without input})  is accurate.}


\textcolor[rgb]{0,0.267,0.588}{\subsection{Comparison between Algorithm \ref{Alg:1} and Algorithm \ref{Alg:kha} }}

Suppose the parameters in system (\ref{equ:sys_without input}) are chosen as
\begin{equation}\label{sys1}
\begin{split}
{\color{V3}\mathcal{A}}= \left[
\begin{array}{cc}
0.8 & 0.2 \\
0.5 & 0.3 \\
\end{array}
\right], \quad\;
{\color{V3}\mathcal{C}}=  \left[
\begin{array}{cc}
1 & 1 \\
\end{array}
\right]
\end{split}
\end{equation}
and the variances of disturbances are selected as $R=0.01$ and $Q={\rm diag}\{0.1,0.1\}$, respectively. In this subsection, let $N_1=N_2=64, M=1024$.

\begin{figure}[H]
	\flushleft
	\includegraphics[width=\linewidth]{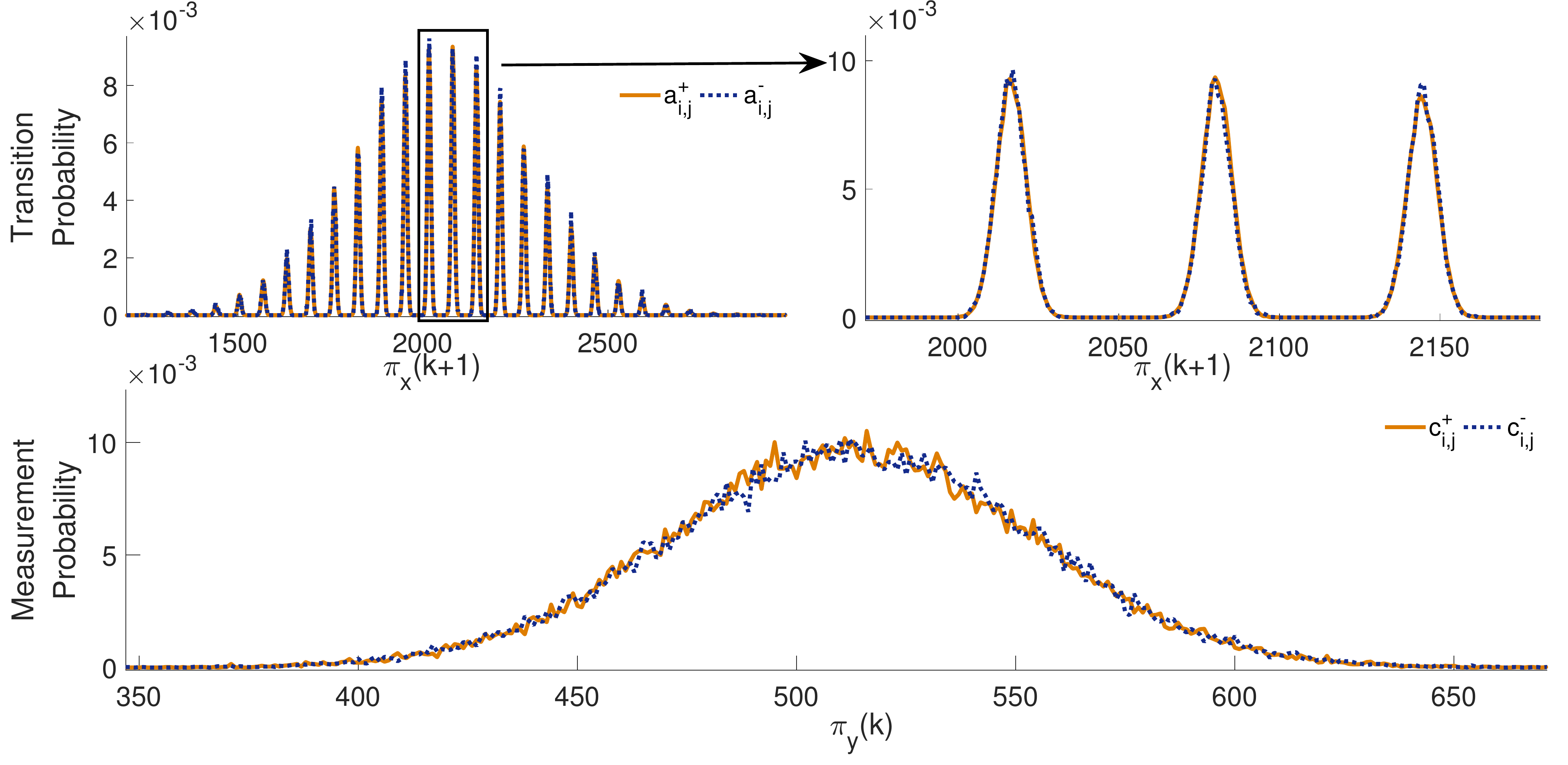}
	\caption{Comparison of a column of the HMM obtained using Algorithm \ref{Alg:1} and Algorithm \ref{Alg:kha}.}\label{fig:colunm}
\end{figure}

{\color{black}The HMM parameters are learned using the naive method in Algorithm \ref{Alg:1} and the improved method in Algorithm \ref{Alg:kha}.  The training results of Algorithm \ref{Alg:1} are noted as $a^+_{i,j}, c^+_{i,j}$ (or $A^+$, $C^+$), while the parameters learnt by Algorithm \ref{Alg:kha} are noted as $a^-_{i,j}, c^-_{i,j}$ (or $A^-$, $C^-$). A comparison of the training results is shown in Fig.~\ref{fig:colunm}.} The small differences between $a^{+}_{i,j}, c^+_{i,j}$ and $a^-_{i,j}, c^-_{i,j}$ illustrate the validity of Algorithm \ref{Alg:kha} since the convergence of Algorithm \ref{Alg:1} is guaranteed by the law of large numbers. 

\textcolor[rgb]{0,0.267,0.588}{\subsection{Applications to Event-triggered Estimation }}
{\color{black}The task of event-triggered state estimation is to estimate the value of state $x(k)$ based on event-triggered measurement information. In this work, the measurement is sent to the estimator according to a pre-specified send-on-delta schedule:}
	\begin{equation}\label{equ:event-tirgger}
	\xi(k)=\left\{\begin{array}{ll}
	0,&\text{$\|y(k)-y(\tau)\|_2<\delta$},\\
	1,&
	\text{otherwise},
	\end{array}\right.
	\end{equation}
where $y(\tau)$ is the previous measurement received by the estimator. 
Since packet dropouts are normally inevitable for a practical communication channel, the remote estimator may not receive the measurements sent out by the sensor (i.e., when $\xi(k)=1$). In this work, the packet dropout effect is modeled by an independent and identically distributed Bernoulli process which is independent of $X(k)$, $Y(k)$ and $\xi(k)$ satisfying
\begin{equation}\label{equ:pack-drop1}
\zeta(k)=\left\{\begin{array}{ll}
0,&\text{there exists a packet loss at time $k$},\\
1,&
\text{there exists no packet loss at time $k$},
\end{array}\right.
\end{equation}
\begin{equation}\label{equ:pack-drop}
\begin{split}
{\rm P}\left(\zeta(k)=1\right) =\lambda,\quad\quad {\rm P}\left(\zeta(k)=0\right)  = 1-\lambda.
\end{split}
\end{equation}
To represent the recipient of the measurement information,  we write
\begin{equation}\label{equ:zonghe}
\gamma(k)=\xi(k)\cdot\zeta(k)=\left\{\begin{array}{ll}
1, &\text{estimator receives $y(k)$},  \\
0, &\text{otherwise}.
\end{array}\right.
\end{equation}
The average communication rate $\eta$ is defined as
\begin{equation}\label{gammab}
\eta:=\frac{1}{N_d}\sum\limits_{k=1}^{N_d}\gamma_k,
\end{equation}
where $N_d$ is the length of the data set.

{\color{black}In this work, three state estimators are compared. The first estimator is Kalman filter with intermittent observations, which is the MMSE state estimator that ignores the unreceived measurement information (see equations (2)-(6) and (13)-(14) in \cite{bs2004}). The second and third estimators are HMM-based state estimators, which are designed using the HMMs obtained by using  Algorithm \ref{Alg:1} and Algorithm \ref{Alg:kha}, respectively (see equations (24) and (54) in \cite{Shi2016Event}).}

To describe the estimation performance of state estimators, we define the average estimation error $E_K, E_{H^{+}}$ and $E_{H^-}$ as
\begin{equation}\label{equ:indexes}
\begin{split}
E_K&=\frac{1}{N_d}\sum\limits_{k=1}^{N_d} \|\hat{x}(k)-x^*(k)\|_2,\\
E_{H^{+}}&=\frac{1}{N_d}\sum\limits_{k=1}^{N_d} \left\|\hat{x}^{+}(k)-x^*(k)\right\|_2,\\
E_{H^{-}}&=\frac{1}{N_d}\sum\limits_{k=1}^{N_d} \left\|\hat{x}^{-}(k)-x^*(k)\right\|_2,
\end{split}
\end{equation}
where $E_{H^{+}}$ is the error of HMM estimator based on $A^{+}$ and $C^+$, $E_{H^{-}}$ is the error of HMM estimator based on $A^{-}$ and $C^{-}$, and $E_K$ is the error of the Kalman filter with intermittent observations, and $x^*$, $\hat x$, $\hat{x}^+$ and $\hat{x}^{-}$ denote the real value, the estimate of the Kalman filter, the estimate using $A^{+}$, $C^+$ and the estimate using $A^{-}$, $C^{-}$, respectively. 

\begin{figure}[htp]
	\flushleft
	\includegraphics[width=\linewidth]{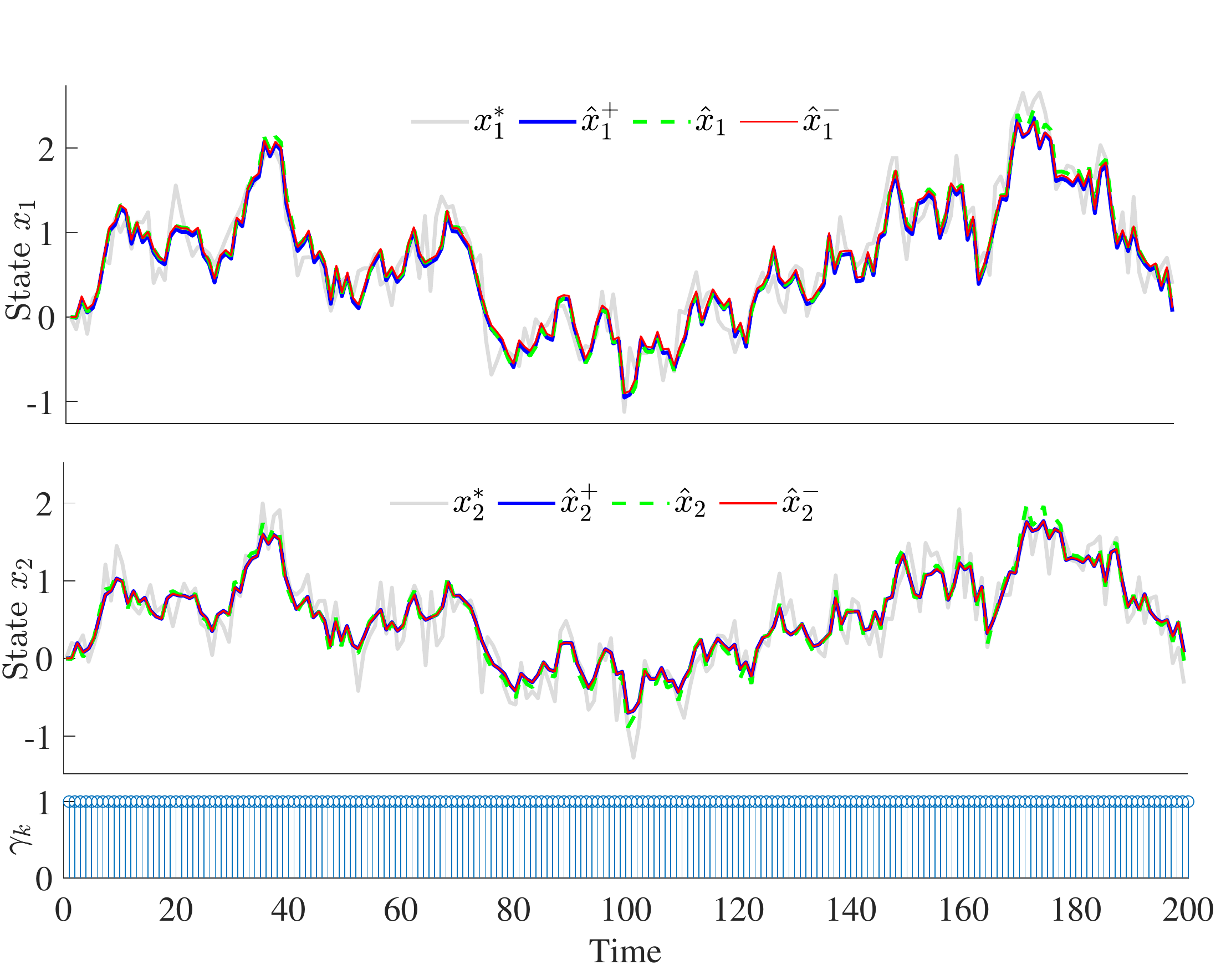}
	\caption{State estimation comparing HMM estimator and Kalman filter, $\eta=1$.}\label{fig:est_com_1}
\end{figure}

{\color{black}First, we consider a limiting case that $\delta$ and $\lambda$ are set to $0$ such that the limiting scenario that the measurements are available for all time instants, to compare the performance of the HMM-based estimators with the standard Kalman filter.  The results are shown in Fig.~\ref{fig:est_com_1}.}
In this case, $E_{H^+}=0.261$, $E_{H^{-}}=0.264$, and $E_K=0.254$, which indicates that HMM-estimators offer good performance for state estimation, nearly as good as the Kalman filter. The consistence observed between $\hat{x}^+$ and $\hat{x}^-$ illustrates the validity of the proposed algorithm (Algorithm~\ref{Alg:kha}).

\begin{figure}[htp]
	\flushleft
	\includegraphics[width=\linewidth]{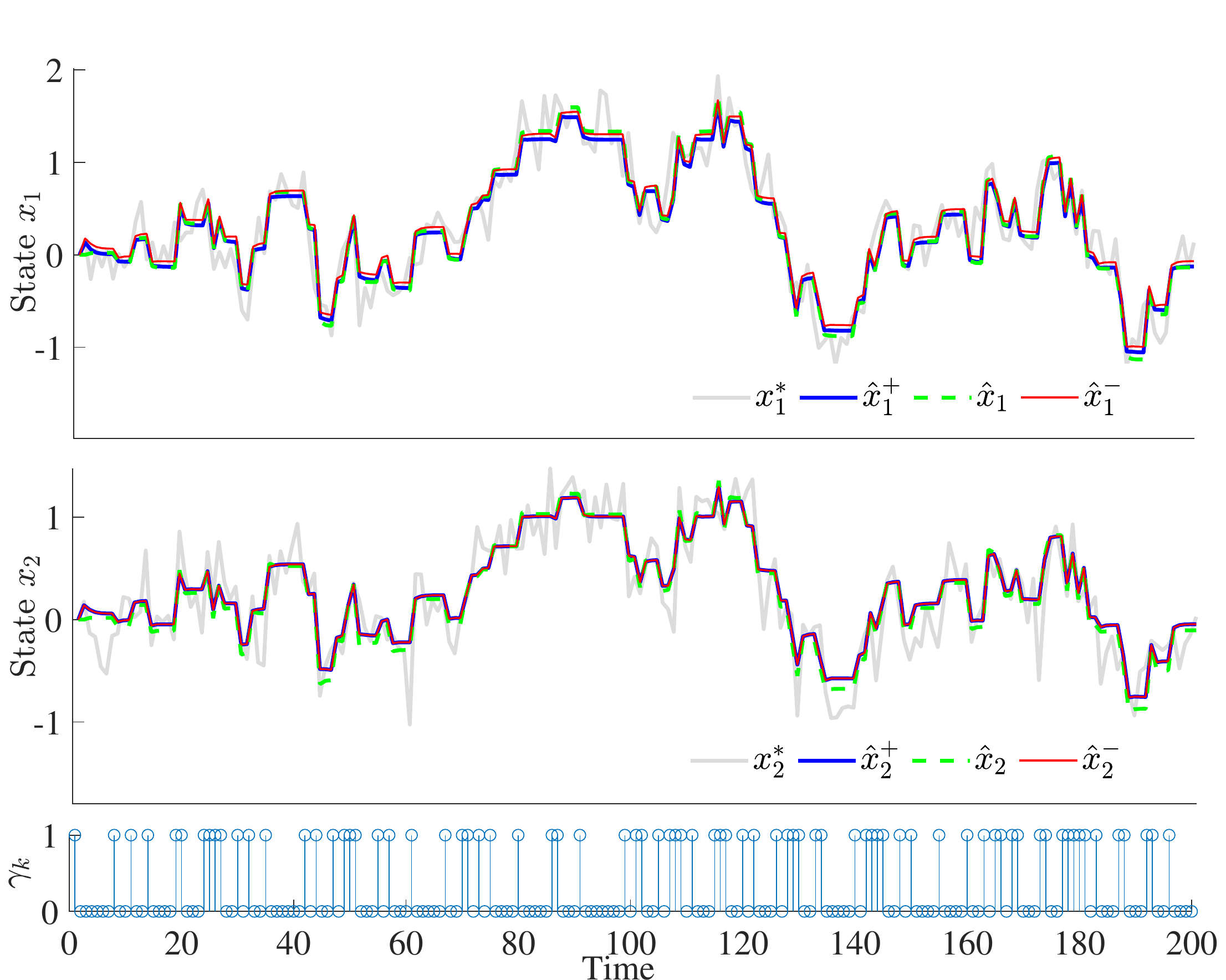}
	\caption{State estimation comparing HMM estimator and Kalman filter, $\delta=0.4081$, $\eta=0.385$.}\label{fig:est_com_3}
\end{figure}

{\color{black}Second, state estimators are applied to event-triggered state estimation through an unreliable channel with $\delta=0.4081$ and $\lambda = 0.95$, and the results are shown in Fig.~\ref{fig:est_com_3} with the average communication rate being $38.5\%$. In Fig.~\ref{fig:est_com_3}, the estimation error $E_{H^+}=0.291$, $E_{H^-} 0.296$, and  $E_K=0.302$. This implies that the HMM-based estimators perform better than the Kalman filter, because of the exploration of the implicit information contained in the event-triggering schedule.}

\begin{figure}[htp]
	\flushleft
	\includegraphics[width=\linewidth]{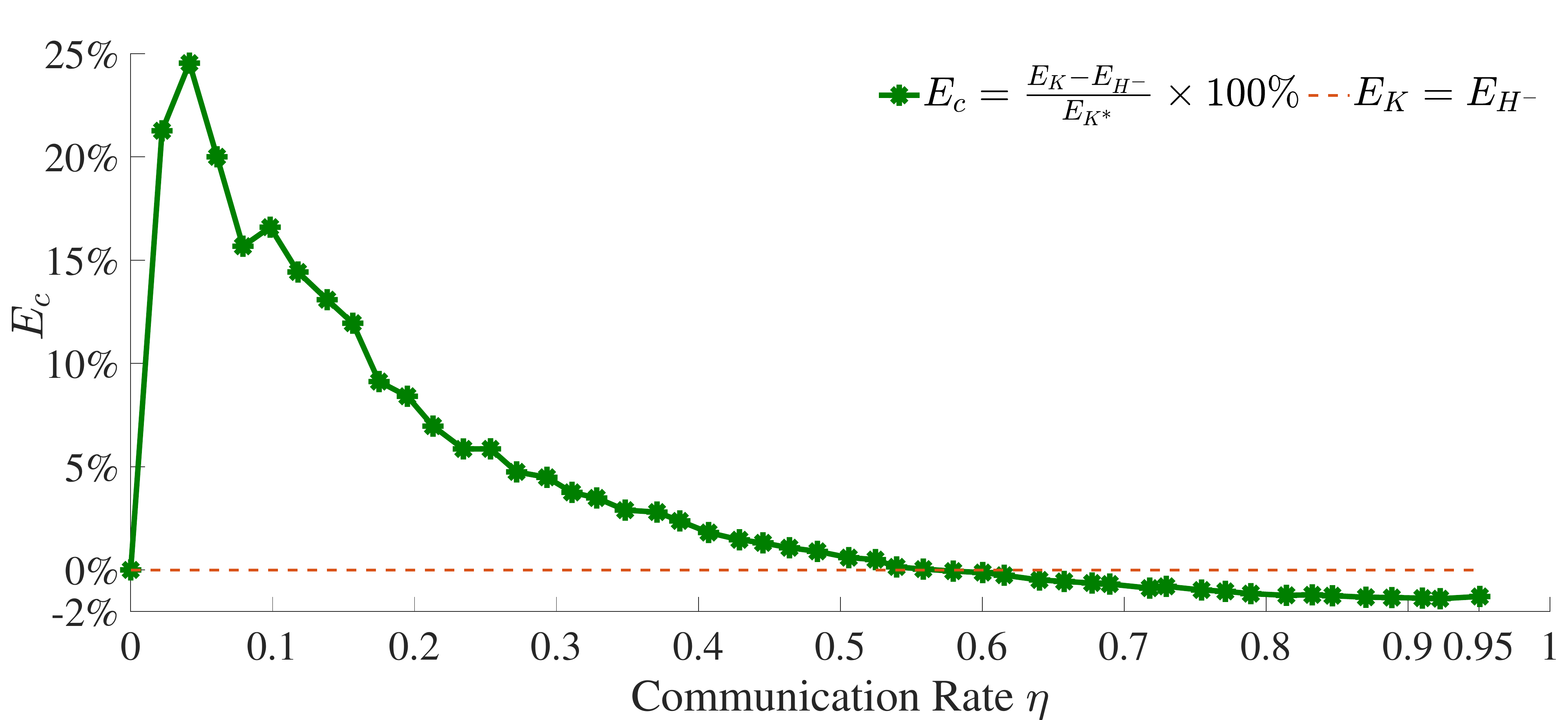}
	\caption{Tradeoff between estimation performance and trigger rate $\eta$ when $\lambda=0.95$ and $\varrho=5$.}\label{tradeoff}
\end{figure}

To further compare the performance of HMM-based estimator using Algorithm~\ref{Alg:kha} and the Kalman filter with intermittent observations, we define
\begin{equation*}
E_c=(E_K-E_{H^-})/E_{K^*},
\end{equation*}
where $E_{K^*}$ is the estimation error of Kalman filter when every measurement is received. 
To evaluate the performance of the event-based estimators, we range the communication rate $\eta$ from $0$ to $1$ by varying the parameter $\delta$ in the event-triggering condition and therefore a sequence of estimation errors for each schedule can be obtained, based on which the tradeoff curves between estimation performance and communication rate are obtained. The results are presented in Fig.~\ref{tradeoff}, which shows that the HMM estimator achieves improved performance in terms of average estimation error compared with Kalman filter with intermittent observations, especially when the communication rate is less than $0.5$. The index $E_c$ is less than zero when the communication rate $\eta$ belongs to $(0.6,0.95)$, which could be caused by the hybrid effect of the  numerical calculation error and the modeling error.

\textcolor[rgb]{0,0.267,0.588}{\section{\textsc{Conclusion}}}\label{conclusion}
{\color{black}In this work, an indirect approach is designed to estimate the parameters of an HMM for a linear Gaussian system. A low-complexity parameter learning algorithm is proposed based on the periodical structural characteristics of the HMM. The convergence of our algorithms is proved using a numerical integration method. Numerical results on model learning and event-triggered state estimation illustrate the motivation and validity of the proposed results. In our next step, the problem of simultaneous HMM modeling and state estimation for a linear Gaussian system will be investigated. }


%

\appendices
\textcolor[rgb]{0,0.267,0.588}{\section{\textsc{Proof of Theorem \ref{thm:ckh}}}}
\begin{pf}
	Considering that the covariance matrix $R$ is a diagonal matrix, the independence between different elements leads to
{\color{V2}	\begin{align}\label{Ptimec}
	&{\rm P}\left(Y(k)=\delta_N^{\pi^y(k)}|X(k)=\delta_N^j\right)\\
	=&\prod\limits_{p=1}^m{\rm P}\left(Y_p(k)=\delta_{M_p}^{\pi_p^y(k)}|X(k)=\delta_N^j\right),
	\end{align} }
	thus we obtain
{\color{V2}	\begin{equation}\label{equ:pf11c}
	c_{\pi^y(k),j}=\prod\limits_{p=1}^m {c}^p_{\pi_p^y(k),j}.
	\end{equation}}
	Noting that
{\color{V2}	\begin{equation*}
	Y_p(k)=\delta_{M_p}^{\pi_p^y(k)},\quad\quad Y(k)=\delta_M^{\pi^y(k)},
	\end{equation*}}
	the combination method \eqref{krony} leads to
{\color{V2}	\begin{equation*}
	\delta_M^{\pi^y(k)}=\delta_{M_1}^{\pi_1^y(k)}\otimes\cdots\otimes\delta_{M_m}^{\pi_n^y(k)}.
	\end{equation*}}
	Thus we obtain
	\begin{equation*}
	[C]^j=[C^1]^j\otimes\cdots\otimes[C^m]^j.
	\end{equation*}
	Then we have
	\begin{equation*}
	C={C}^1*{C}^2*\cdots*{C}^m
	\end{equation*}
	based on the definition of Khatri-Rao product, which completes the proof. \hfill\QEDclosed
\end{pf}
\textcolor[rgb]{0,0.267,0.588}{\section{\textsc{Proof of Theorem \ref{colcij}}}}\label{appendixc}
\begin{pf}
{\color{V2}	Based on Lemma \ref{thm:cijt}, we obtain
	\begin{footnotesize}
		\begin{align}	
		\hat{c}^p_{i_{\rm\textsc t},j_{\rm\textsc t}}=&\left(\overline{\alpha}_p^y(k;{\rm\textsc t})-\underline{\alpha}_p^y(k;{\rm\textsc t})\right)(\overline{y}_p(k;{\rm\textsc t})-\underline{y}_p(k;{\rm\textsc t}))\notag\\
		\quad&f_p^y\left(\mu_{\rm\textsc t}^y,\frac{\overline{y}_p(k;{\rm\textsc t})+\underline{y}_p(k;{\rm\textsc t})}{2}-\mu_{\rm\textsc t}^y\right);\label{zm1}\\
		\hat{c}^p_{i_{\rm\textsc s},j_{\rm\textsc s}}=&\left(\overline{\alpha}_p^y(k;{\rm\textsc s})-\underline{\alpha}_p^y(k;{\rm\textsc s})\right)(\overline{y}_p(k;{\rm\textsc s})-\underline{y}_p(k;{\rm\textsc s}))\notag\\
		\quad&f_p^y\left(\mu_{\rm\textsc s}^y,\frac{\overline{y}_p(k;{\rm\textsc s})+\underline{y}_p(k;{\rm\textsc s})}{2}-\mu_{\rm\textsc s}^y\right).\label{zm2}	
		\end{align}
	\end{footnotesize}
	According to {\color{V3-2}equation} \eqref{fpy}, we have
	\begin{small}
		\begin{align}
		&f_p^y\left(\mu_{\rm\textsc t}^y,\frac{\overline{y}_p(k ;{\rm\textsc t})+\underline{y}_p(k ;{\rm\textsc t})}{2}-\mu_{\rm\textsc t}^y\right)\notag\\
		=&
		f_{\rm G}\left(\mu_{\rm\textsc t}^y,\bar{\sigma}_p^y\right)
		\cdot f_{\rm G}\left(\frac{\overline{y}_p(k ;{\rm\textsc t})+\underline{y}_p(k ;{\rm\textsc t})}{2}-\mu_{\rm\textsc t}^y,r_p\!\right);\label{zm3}\\
		&f_p^y\left(\mu_{\rm\textsc s}^y,\frac{\overline{y}_p(k ;{\rm\textsc s})+\underline{y}_p(k ;{\rm\textsc s})}{2}-\mu_{\rm\textsc s}^y\right)\notag\\
		=&
		f_{\rm G}\left(\mu_{\rm\textsc s}^y,\bar{\sigma}_p^y\right)
		\cdot f_{\rm G}\left(\frac{\overline{y}_p(k ;{\rm\textsc s})+\underline{y}_p(k ;{\rm\textsc s})}{2}-\mu_{\rm\textsc s}^y,r_p\!\right).\label{zm4}
		\end{align}
	\end{small}

	Combining equations \eqref{zm1} and \eqref{zm3}, \eqref{zm2} and \eqref{zm4}, we write
	\begin{align}
	\hat{c}^p_{i_{\rm\textsc t},j_{\rm\textsc t}}&\!=\!\tau_p^y(k;{\rm\textsc t})f_{\rm G}\left(\frac{\overline{y}_p(k;{\rm\textsc t})\!+\!\underline{y}_p(k;{\rm\textsc t})}{2}\!-\!\mu_{\rm\textsc t}^y,r_p\!\!\right);\label{cpt}\\
	\hat{c}^p_{i_{\rm\textsc s},j_{\rm\textsc s}}&\!=\!\tau_p^y(k;\!{\rm\textsc s})f_{\rm G}\left(\frac{\overline{y}_p(k;{\rm\textsc s})\!+\!\underline{y}_p(k;{\rm\textsc s})}{2}\!-\!\mu_{\rm\textsc s}^x,r_p\!\!\right)\!,\label{cps}
	\end{align}
	where $\tau_p^y(k;{\rm\textsc t})$ and $\tau_p^y(k;{\rm\textsc s})$ are constants as
	{\small
		\begin{align*}
		\tau_p^y(k;{\rm\textsc t})&\!=\!\left(\!\overline{\alpha}_p^y(k;\!{\rm\textsc t})\!-\!\underline{\alpha}_p^y(k;\!{\rm\textsc t})\!\right)\!(\overline{y}_p(k;{\rm\textsc t})\!-\!\underline{y}_p(k;{\rm\textsc t})\!)f_{\rm G}\left(\mu_{\rm\textsc t}^y,\!\bar{\sigma}_p^y\right)\\
		\tau_p^y(k;{\rm\textsc s})&\!=\!\left(\!\overline{\alpha}_p^y(k;\!{\rm\textsc s})\!-\!\underline{\alpha}_p^y(k;\!{\rm\textsc s})\!\right)\!\!(\overline{y}_p(k;{\rm\textsc s})\!-\!\underline{y}_p(k;{\rm\textsc s})\!)f_{\rm G}\!\left(\mu_{\rm\textsc s}^y,\!\bar{\sigma}_p^y\right).
		\end{align*}}
	Equations \eqref{cpt} and \eqref{cps} show that, a transition probability of the target column $\hat{c}^p_{i_{\rm\textsc t},j_{\rm\textsc t}}$ can be seen as a product of a constant $\tau_p^y(k;{\rm\textsc t})$ (called constant part) and a Gaussian distribution function with variance $r_p^2$ (called Gaussian part), while a transition probability of the standard column $\hat{c}^p_{i_{\rm\textsc s},j_{\rm\textsc s}}$ can be seen as a product of a constant $\tau_p^y(k;{\rm\textsc s})$ (constant part) and a Gaussian distribution function with variance $r_p^2$ (Gaussian part). Thus a constant $\kappa_y$ is formed as $\kappa_y=\frac{\tau_p^y(k;{\rm\textsc t})}{\tau_p^y(k;{\rm\textsc s})}$ to close the gap between the constant parts of different columns.
	
	Then, for Gaussian parts of $\hat{c}^p_{i_{\rm\textsc s},j_{\rm\textsc s}}$ and $\hat{c}^p_{i_{\rm\textsc t},j_{\rm\textsc t}}$, consider the elements of the standard column and the target column in a same row, where $i_{\rm\textsc t}=i_{\rm\textsc s}$, $\overline{y}_p(k;{\rm\textsc t})=\overline{y}_p(k;{\rm\textsc s})$, $\underline{y}_p(k;{\rm\textsc t})=\underline{y}_p(k;{\rm\textsc s})$. It is obvious that
	\begin{align}\label{utuy}
	&f_{\rm G}\left(\frac{\overline{y}_p(k;{\rm\textsc t})+\underline{y}_p(k;{\rm\textsc t})}{2}\!-\!\mu_{\rm\textsc t}^y,r_p\!\right)\\
	=&f_{\rm G}\left(\frac{\overline{y}_p(k;{\rm\textsc s})+\underline{y}_p(k;{\rm\textsc s})}{2}\!-\!\mu_{\rm\textsc s}^y-(\mu_{\rm\textsc t}^y-\mu_{\rm\textsc s}^y),r_p\!\right).\notag
	\end{align}
	For quantized transition probabilities, $(\mu_{\rm\textsc t}^y-\mu_{\rm\textsc s}^y)$ needs to be quantized to a integer multiples of $h_p^y$, where $h_p^y=\overline{y}_p(k)-\underline{y}_p(k)$ is the length of quantization sub-intervals in \eqref{quany}. Therefore, $(\mu_{\rm\textsc t}^y-\mu_{\rm\textsc s}^y)$ is quantized as $g_p^yh_p^y$, where $g_p^y=\left\lfloor\frac{\mu_{\rm\textsc t}^y-\mu_{\rm\textsc s}^y}{h_p^y} \right\rfloor$.
	Since
	\begin{align*}
	&\left|(\mu_{\rm\textsc t}^y-\mu_{\rm\textsc s}^y)-g_p^yh_p^y\right|
	=\left|(\mu_{\rm\textsc t}^y-\mu_{\rm\textsc s}^y)-\left\lfloor\frac{\mu_{\rm\textsc t}^y-\mu_{\rm\textsc s}^y}{h_p^y} \right\rfloor h_p^y\right|
	\leq h_p^y,
	\end{align*}
	we obtain
	\begin{align*}
	&\lim\limits_{h_p^y\rightarrow 0}|(\mu_{\rm\textsc t}^y-\mu_{\rm\textsc s}^y)-g_p^yh_p^y|=0,\\
	&\lim\limits_{h_p^y\rightarrow 0}\left|\left[\frac{\overline{y}_p(k;{\rm\textsc s})\!+\!\underline{y}_p(k;{\rm\textsc s})}{2}\!-\!\mu_{\rm\textsc s}^y-(\mu_{\rm\textsc t}^y-\mu_{\rm\textsc s}^y)\right]\right.\\
	&\quad-\left.\left[\frac{\overline{y}_p(k;{\rm\textsc s})+\underline{y}_p(k;{\rm\textsc s})}{2}\!-\!\mu_{\rm\textsc s}^y-g_p^yh_p^y\right]\right|=0.
	\end{align*}
	Then the uniform continuity of the Gaussian distribution function leads to
	\begin{small}
		\begin{align}
		&\lim\limits_{h_p^y\rightarrow 0}\left|f_{\rm G}\!\left(\frac{\overline{y}_p(k;{\rm\textsc s})\!+\!\underline{y}_p(k;{\rm\textsc s})}{2}\!-\!\mu_{\rm\textsc s}^y-(\mu_{\rm\textsc t}^y-\mu_{\rm\textsc s}^y),r_p\!\right)\right.\notag\\
		&\;\left. -f_{\rm G}\!\left(\frac{\overline{y}_p(k;{\rm\textsc s})+\underline{y}_p(k;{\rm\textsc s})}{2}\!-\!\mu_s^y-g_p^yh_p^y,r_p\right)\right|=0.
		\end{align}
	\end{small}
Note that
\begin{footnotesize}
	\begin{align}
	&\hat{c}^p_{i_{\rm\textsc t},j_{\rm\textsc t}}\label{cptk}\\
	=&\tau_p^y(k;{\rm\textsc t})f_{\rm G}\left(\frac{\overline{y}_p(k;{\rm\textsc t})\!+\!\underline{y}_p(k;{\rm\textsc t})}{2}\!-\!\mu_{\rm\textsc t}^y,r_p\!\!\right);\notag\\
	&\kappa_y\hat{c}^p_{i_{\rm\textsc s}-g_p^y,j_{\rm\textsc s}}\label{tcpsk}\\
	=&\frac{\tau_p^y(k;{\rm\textsc t})}{\tau_p^y(k;{\rm\textsc s})}\tau_p^y(k;\!{\rm\textsc s})f_{\rm G}\left(\frac{\overline{y}_p(k;{\rm\textsc s})\!+\!\underline{y}_p(k;{\rm\textsc s})}{2}\!-\!\mu_{\rm\textsc s}^y\!-\!g_p^yh_p^y,r_p\!\!\right).\notag
	\end{align}
\end{footnotesize}
Equation \eqref{limyth} can be obtained by combining equations \eqref{utuy}-\eqref{tcpsk}. Thus the claim holds.}	\hfill\QEDclosed
\end{pf}



\ifCLASSOPTIONcaptionsoff
\newpage
\fi



\footnotesize
\bibliographystyle{IEEEtran}
\bibliography{kk7}
%



%
 






\end{document}